\documentclass[onecolumn,aps,prd,preprintnumbers,showpacs,superscriptaddress,nofootinbib,amsmath,amssymb,floats,floatfix,showkeys,notitlepage,longbibliography]{revtex4-2}
\pdfoutput=1
\usepackage{orcidlink}
\usepackage{lipsum}
\usepackage{graphicx}
\usepackage{subfigure}
\usepackage{braket}
\usepackage[utf8]{inputenc}
\usepackage{sans}
\usepackage{changes}
\usepackage{hyperref}
\hypersetup{colorlinks=true,linkcolor=blue,urlcolor=blue,citecolor=blue}
\usepackage[toc,page]{appendix}
\usepackage[normalem]{ulem}
\usepackage{adjustbox}
\usepackage{latexsym}
\usepackage{amsmath}
\usepackage{amssymb}
\usepackage{amsfonts}
\usepackage{dcolumn}
\usepackage{bm}
\usepackage{tikz}
\usepackage{bigints}
\usepackage{comment}
\usepackage{array,tabularx,multirow,booktabs}
\usepackage[tracking=true]{microtype}
\SetTracking{}{500}
\SetTracking{encoding={*}, shape=sc}{40}
\UseRawInputEncoding 
\allowdisplaybreaks

\newcommand{\galla}{\lambda}
\begin{document} \sloppy

\title{ Dynamics of Interacting Murnaghan-Equation-of-State Fluids with Global Monopole and Nonlinear Power-Yang-Mills Hair in Critical-Dimension AdS Black Holes}

\author{Y. Sekhmani}
\email[Email: ]{sekhmaniyassine@gmail.com}
\affiliation{Center for Theoretical Physics, Khazar University, 41 Mehseti Street, Baku, AZ1096, Azerbaijan.}
\affiliation{Centre for Research Impact \& Outcome, Chitkara University Institute of Engineering and Technology, Chitkara University, Rajpura, 140401, Punjab, India}

\author{S.K. Maurya}
\email[Email: ]{sunil@unizwa.edu.om}
\affiliation{Department of Mathematical and Physical Sciences, College of Arts and Sciences, University of Nizwa, P.O. Box 33, Nizwa 616, Sultanate of Oman}

\author{J. Rayimbaev}
\email[Email: ]{javlon@astrin.uz}
\affiliation{
Tashkent International University of Education, Imom Bukhoriy 6, Tashkent 100207, Uzbekistan}
\affiliation{University of Tashkent for Applied Sciences, Gavhar Str. 1, Tashkent 100149, Uzbekistan}
\affiliation{
National University of Uzbekistan, Tashkent 100174, Uzbekistan}
\affiliation{Tashkent State Technical University, Tashkent 100095, Uzbekistan}
\affiliation{Urgench State University, Kh. Alimjan Str. 14, Urgench 221100, Uzbekistan}

\author{M. Altanji}
\email[Email: ]{maltenji@kku.edu.sa}
\affiliation{Department of Mathematics, College of Sciences, King Khalid University, Abha 61413, Saudi Arabia}

\author{I. Ibragimov} 
\email[Email: ]{i.ibragimov@kiut.uz}
\affiliation{Kimyo International University in Tashkent, Shota Rustaveli street 156, Tashkent 100121, Uzbekistan}

 \author{A. Landry}
\email[Email: ]{a.landry@dal.ca}
\affiliation{Department of Mathematics and Statistics, Dalhousie University, Halifax, Nova Scotia, Canada, B3H 3J5}

\author{S. Muminov}
\email[Email: ]{sokhibjan.muminov@mamunedu.uz}
\affiliation{Mamun University, Bolkhovuz Street 2, Khiva 220900, Uzbekistan}

\date{\today}
\begin{abstract}
We construct and analyze a novel family of exact higher-dimensional black hole (BH) solutions in Einstein-Power-Yang-Mills (PYM) gravity minimally coupled to a scalar field (SF) multiplet supporting a global monopole (GM) and a surrounding anisotropic “scalar gas” whose stress-energy obeys a generalized Murnaghan (dual-polytropic) equation of state (EoS). By adopting the Wu--Yang (WY) magnetic ansatz for the non-Abelian sector and enforcing the physically motivated radial condition $p_r=-\rho$, the field equations (FEs) admit closed-form expressions for the matter density and a single-function lapse $F(r)$ expressed in terms of elementary and Gauss hypergeometric functions. The Murnaghan fluid interpolates between a non-linear core and an effective vacuum at a large radius, producing a backreaction encoded by parameters $\beta$, $\gamma$, $\xi$, and $\chi$ that control the stiffness and scalar-backreaction scale. We perform a systematic survey of classical energy conditions (EC), identifying regions of parameter space where the null energy condition/dominant energy condition (NEC/DEC) holds while the strong energy condition (SEC) is generically violated (phantom-like behavior), and examine curvature invariants to demonstrate that the solutions possess a central curvature singularity for $n>3$. Thermodynamic properties are derived in the extended phase space: explicit formulas for the Hawking temperature $T(r_h)$, entropy $S$, conjugate potentials, and a generalized Smarr relation are obtained; the heat capacity $C_{P,Q}$ exhibits divergencies and sign changes that mark local stability boundaries and second-order phase transitions. Varying the Yang-Mills (YM) charge $Q$, the nonlinearity index $\epsilon$, the monopole coupling $\eta$, and the Murnaghan parameters generates a rich phase behavior structure and topology changes in the defect ($\varphi$-Duan) map. The results underscore the manner in which gauge nonlinearity, topological defects, and dual-polytropic matter collaboratively transform BH thermodynamics and the topology of phase spaces.
\end{abstract}

\keywords{Murnaghan EoS, Monopole Fluid, Power-Yang-Mills fields, Black Hole, Duan-$\varphi$ map.}

\maketitle

\section{Introduction} \label{intro}

General relativity (GR) remains the preeminent classical theory of gravitation, furnishing an extraordinarily successful description of phenomena ranging from planetary dynamics to the behavior of compact objects. However, its predictive power is known to fail in certain extreme regimes \cite{Penrose:1964wq,Hawking:1970zqf}: the singularity theorems of Penrose and Hawking show that, within the classical GR framework, curvature singularities and the attendant loss of determinism are generic under rather mild assumptions \cite{Rovelli:2004tv,Brunnemann:2005in,Modesto:2007wp,Bodendorfer:2011nx,Long:2019nkf,Long:2020wuj,Bojowald:2001xe}. This breakdown motivates two complementary avenues of investigation: the formulation of modified classical gravitational theories (for example, those that incorporate spacetime torsion as in the Einstein--Cartan framework \cite{Hehl:1976kj}) and the development of quantum--gravity (QG) paradigms (such as string theory and loop QG) that provide mechanisms for ultraviolet completion and singularity resolution \cite{Green:1982tc,Han:2005km,Ashtekar:2011ni,Zhang:2022vsl,Gambini:2008dy}. In the context of this broader field, theories coupling gravity to non-Abelian gauge fields occupy a privileged position. The Einstein-Yang-Mills (EYM) system, which embeds Yang-Mills (YM) fields into the geometrical structure of curved spacetime, provides both a natural arena to probe gauge gravity interactions and a fertile source of novel compact object solutions \cite{Jackiw:1976pf,Brink:1976bc,Dorigoni:2024vrb,Koutrolikos:2024neu}. Beyond model construction, EYM systems have been instrumental in clarifying issues of asymptotic symmetry, black hole (BH) structure and stability, as well as aspects of semiclassical and quantum gravitational physics \cite{Barnich:2013sxa}.

BHs provide one of the most incisive observational and theoretical testing grounds for gravitational theories. Recent observational breakthroughs notably the Event Horizon Telescope (EHT) images of M87* (2019) and Sgr A* (2022)  offer horizon scale probes of photon dynamics in the strong field regime and have opened an empirical window onto the structure of spacetime near compact objects \cite{EventHorizonTelescope:2019dse,EventHorizonTelescope:2019uob,EventHorizonTelescope:2019jan,EventHorizonTelescope:2019ths,EventHorizonTelescope:2019pgp,EventHorizonTelescope:2019ggy,EventHorizonTelescope:2022wkp}. The characteristic "shadow" captured by the EHT corresponds to the critical curve determined by unstable spherical photon orbits; its morphology and angular scale are tightly linked to the underlying spacetime metric and therefore constitute sensitive diagnostics for competing theories of gravity \cite{Guo:2020qwk,Johannsen:2013vgc,Synge:1966okc,Johannsen:2010ru,Li:2021riw, Sui:2023tje, Abdujabbarov:2016hnw, Wei:2013kza,Mizuno:2018lxz,Konoplya:2019sns,Zare:2024dtf,Sekhmani:2025bsi,Al-Badawi:2025gvx,Fathi:2025ikx,Sekhmani:2024dhc,Al-Badawi:2025coy,Sekhmani:2025xzf}. In addition to this conceptual development, the theoretical investigations, ranging from particle dynamics and quasi-periodic oscillations to gravitational lensing and the motion of spinning particles \cite{QPo1,QPo2,QPo3}, serve not only to constrain the key parameters of the BH but also to enhance our understanding of its physical nature.

Motivated by both theoretical considerations and observational opportunity, there has been growing interest in BH solutions sourced by non-linear electromagnetic sectors, in particular those governed by powers of the Maxwell invariant $(F_{\mu\nu}F^{\mu\nu})^\epsilon$. This class of models generalizes the linear Maxwell case ($\epsilon=1$) and is qualitatively distinct from other non-linear frameworks such as Born-Infeld electrodynamics; an explicit Taylor expansion of the Born-Infeld square root Lagrangian around small field strength makes these structural differences manifest \cite{Born:1934,Dyadichev:2000xh,Alam:2021}. Of special significance is the choice $\epsilon=N/4$, where $N$ denotes the spacetime dimension: for this value, the EM tensor of the power-law (PL) electromagnetic field is traceless, and the electromagnetic sector attains conformal invariance, which in turn enforces a vanishing scalar curvature in the absence of additional fields or a cosmological constant \cite{Hassaine:2007,Sheykhi:2012}. Such conformal properties are particularly attractive in higher-dimensional settings and have important consequences for the existence and structure of BH solutions and their thermodynamics \cite{Hassaine:2007,Dehghani:2016}. 

Analytic higher-dimensional constructions exploiting the Wu--Yang (WY) ansatz produce tractable, purely magnetic configurations that render the field equations (FEs) solvable and expose the explicit $r$-dependence induced by the gauge sector \cite{Yasskin:1975,Mazharimousavi:2008}. Extending these ideas, PYM sources and polynomial YM hierarchies:
\[
\mathcal{L}_{\rm YM}^{\rm (generalised)}=\sum_{k=0}^{P} b_k\big(F^{(a)}_{\mu\nu}F^{(a)\,\mu\nu}\big)^k
\]
have been studied in Einstein, Lovelock and quasi-topological gravities, where the nonlinearity parameter $\epsilon$ (or the set $\{b_k\}$) generates novel $r$-dependent terms in the metric function and modifies horizon structure and thermodynamics relative to canonical EYM solutions \cite{Mazharimousavi:2009,Ali:2020}.  Physical viability further constrains admissible $\epsilon$: negative $\epsilon$ often violates standard energy and causality conditions. In contrast, the set of integer (or rational) $\epsilon$ compatible with the energy conditions (ECs) depends sensitively on $N$ (for instance, analyses show that for low dimensions the canonical $\epsilon=1$ is often singled out whereas for higher $N$ other positive integers/rationals become admissible) \cite{Gurtug:2012,Zhang:2015,Stetsko:2020}.  These changes also leave their imprint on the thermodynamic phase structure: power-gauge BHs admit modified Smarr relations, enriched $P\!-\!V$ criticality and, in certain non-linear branches, reentrant or zeroth-order transitions absent in the linear case \cite{MasoumiJahromi:2023,Ali:2020}.  The non-Abelian generalization we adopt here is a a higher-dimensional WY magnetic ansatz coupled to a PYM/polynomial YM Lagrangian, therefore provides a controlled laboratory to study how finite nonlinearity gauge sectors and EC constraints reshape both local (near-horizon) geometry and global thermodynamic behavior.

Topological defects such as global monopoles (GMs) further enrich the spectrum of BH geometries. GMs, which arise from the spontaneous breaking of a global $\mathrm{SO}(3)$ symmetry, imprint a solid-angle deficit on the ambient spacetime and thereby modify geodesic structure, light deflection and horizon properties \cite{10,11,12,13}. Both ordinary GMs (OGM), endowed with canonical (positive) kinetic energy, and phantom GMs (PGM), characterized by negative kinetic terms that violate standard ECs, have been extensively investigated; their embedding in the BH backgrounds provides a versatile theoretical laboratory to probe departures from the predictions of GR, with observable consequences for causal structure, lensing, thermodynamics and quasinormal-mode spectra \cite{14,15}.

The Murnaghan EoS, originally introduced in solid state physics to describe the pressure--volume response of crystalline media under compression~\cite{Murnaghan:1944}, has recently been adopted as a flexible phenomenological model in cosmology and gravitation~\cite{Dunsby:2024,Luongo:2025,Muccino:2025}. Interpreted as a superposition of two polytropic components, the Murnaghan EoS provides a versatile description of effective fluids that can mimic a range of cosmological constituents; this dual–polytropic viewpoint is particularly useful for modelling the combined, interacting effects of dark energy (DE) and dark matter (DM) and, in appropriate limits, for reproducing the behavior of logotropic and generalized Chaplygin-type fluids while remaining close to the conventional $\Lambda$CDM expansion history~\cite{Dunsby:2024,Luongo:2025,Muccino:2025}. 

In this work we adopt the Murnaghan ansatz
\begin{equation}
P(\rho) = -\beta\,\rho - \galla\,\rho^{1+\frac{1}{\xi}} \,,\label{eq:MurnaghanEoS}
\end{equation}
with $\mu,\lambda,\xi$ constant parameters controlling the linear and non-linear contributions. This form can be regarded as a non-linear generalization of the standard polytropic law $P\propto\rho^{1+1/n}$: in the infinitesimal–strain regime $\Delta\rho/\rho_0\ll1$ it reduces to a pure PL, while—thanks to Murnaghan's phenomenological choice $K(P)=K_0+K_0'P$ for the bulk modulus—it remains accurate for finite density variations up to $\Delta\rho/\rho_0\sim\mathcal{O}(1)$ without invoking higher–order corrections~\cite{Murnaghan:1944,Wallace:1972,Birch:1947,Weppner:2015}.

Several useful properties follow immediately from \eqref{eq:MurnaghanEoS}. The fluid's adiabatic sound speed (squared) is given by
\begin{equation}\label{eq:soundspeed}
c_s^2(\rho) \;=\; \frac{dP}{d\rho}
\;=\; -\mu \;-\; \lambda\Big(1+\tfrac{1}{\xi}\Big)\rho^{\,\frac{1}{\xi}} .
\end{equation}
Physical stability and causality then place constraints on the parameter space: for linearised perturbations one typically requires $0\leq c_s^2\leq 1$, which restricts the allowed combinations of $\mu,\lambda,\xi$ for the density range of interest. In particular, for $\lambda>0$ the non-linear term makes $c_s^2$ increasingly negative (or positive) with $\rho$ depending on the sign of $\lambda$ and the value of $\xi$; this behavior must be checked against ECs and observational priors when fitting the model to astrophysical or cosmological data.

Previous studies have shown that such an ansatz can produce regular BH solutions with non-trivial thermodynamic behavior, including phase transitions and intricate stability properties \cite{Dymnikova:1992,Balart:2014,Ansoldi:2008,Culetu:2011,Bronnikov:2001,Mazharimousavi:2009,Kumar:2020,Meitei:2022,Sekhmani:2025kav,Sekhmani:2025fpw,Sekhmani:2024kfj,Sekhmani:2024udl,Sekhmani:2023plr}. In this paper we synthesise these elements by constructing and analysing higher‑dimensional BH solutions sourced by PL YM fields in the presence of GM defects and a surrounding Murnaghan fluid. We characterize the resulting geometries, delineate the parameter regimes compatible with standard energy and causality conditions, and explore the thermodynamic consequences induced by the Murnaghan EoS. Our objective is to elucidate how the interplay between gauge sector non-linearity, topological defects, and a dual polytropic matter model modifies horizon structure, and the thermodynamic phase space of BHs, thereby providing new theoretical diagnostics to confront alternative gravitational scenarios with observations.

In this work, we build on these developments to explore a BH solution in higher-dimensional spacetime within the framework of Einstein (–$\Lambda$)--PYM theory. The gravitational configuration is further examined in the presence of a surrounding scalar gas governed by the Murnaghan EoS, together with a topological defect generated by a GM. Building upon this setup, the paper is structured as follows: Section \ref{sec1} introduces the theoretical framework, including the Ricci scalar, the PYM invariant, the GM parametrised by a real scalar field (SF), and the Murnaghan fluid with negative pressure, which effectively mimics DE. Section \ref{sec2} is devoted to solving the FEs, yielding an exact analytical expression for the scalar gas energy density, together with an examination of the ECs. This, in turn, leads to the corresponding gravitational metric function and a detailed analysis of the associated curvature singularities. To establish the physical viability of the derived BH solution, Section \ref{sec3} provides a comprehensive analysis of its thermodynamic properties. Therefore, Section \ref{sec4} interprets the thermodynamic findings from a topological perspective by employing the $\varphi$--Duan map, with particular emphasis on the role of spacetime dimensionality in the presence of a second-order phase transition. Finally, Section \ref{sec5} summarizes the main findings and outlines future challenges associated with extending the framework of the DE--mimicking fluid.

\section{Scalar Gas Model of Dark Energy with Murnaghan EoS: $p =-(\beta+\gamma \,\rho^{1+\frac{1}{\xi}})$}
\label{sec1}

To establish a consistent physical framework for charged scalar gas fields within GR, it is essential to construct an appropriate action principle. To this end, we consider a higher-dimensional Einstein-PYM theory with a cosmological constant, minimally coupled to a SF multiplet $\phi^a$ that represents a GM. The matter sector is further characterised by a Murnaghan-type EoS, with the full action expressed as follows:
    \begin{equation}\label{action}
    \mathcal{I}=\int d^nx\sqrt{-g}\left[\frac{1}{2\kappa}(\mathcal{R}-2\Lambda )-\mathcal{F}^\epsilon+\mathcal{L}[\phi^a]\right]+\mathcal{I}_{\text{Mur}},
\end{equation}
where 
$\Lambda=−(n-1)(n-2)/
2L^2$ is the cosmological constant with $L$ is the Anti-de Sitter (AdS) radius\footnote{In the following, we assume that $\kappa=8\pi G=c=1$, where $G$ represents the Newtonian gravitational constant and $c$ denotes the speed of light.}, $g=\det (g_{\mu\nu})$ is the determinant of the metric 2-rank tensor $g_{\mu\nu}$, $\mathcal{I}_{Mur}$ stands for matter with Murnghan EoS and $\mathcal{F}$ is the YM invariant defined as
\begin{equation}
\mathcal{F}=\mathrm{h}_{ij}F_{\mu\nu}^{(i)}F^{(j)\mu\nu},\quad\mathrm{h}_{ij}\equiv-\frac{\mathcal{H}_{(i)(j)}}{\mid\mathcal{H}^{1/N}\mid}.
\end{equation}
Here, $\mathcal{H}_{(i)(j)}= C_{(i)(l)}^{(k)}C_{(j)(k)}^{(l)}$ is the metric tensor of the gauge group, and $\text{det}\mathcal{H}_{ij}$ is the related determinant. The indices $i$, $j$, and $k$ take values from 1 to $N$, $C_{jk}^i$ are the structure constants of the gauge group theory, and the power $\epsilon$ is selected as an arbitrary positive real number, as choosing $\epsilon < 0$ would result in the violation of ECs and causality conditions. In accordance with this notation, the gauge field tensor can be expressed by:
\begin{equation}
    F_{\mu\nu}^{(i) }=\partial_\mu A_\nu^{(i)}-\partial_\nu A_\mu^{(i)}+\frac{1}{2\sigma}C_{(j)(k)}^{(i)}A_\mu^{(j)}A_{\nu}^{(k)}.
\end{equation}
where $\sigma$ represents the coupling constant of the non-abelian theory, while $A_\mu^{(i)}$ denotes the gauge potentials.

The Lagrangian describing the SF is given by:
\begin{equation}
  \mathcal{L}
    = -\tfrac{1}{2}\bigl(\partial_\mu\phi^a\bigr)\bigl(\partial^\mu\phi^a\bigr)
      \;-\; \mathcal{V}(\phi^a)\,,  
  \qquad
  \mathcal{V}(\phi^a)
    = \frac{\lambda}{4}\bigl(\phi^a\phi^a - \eta^2\bigr)^2\,,  
\end{equation}
where $a=1,\dots,n-1$.  Here $\eta$ sets the scale of spontaneous breaking of the global $O(n-1)$ symmetry down to $O(n-2)$\,\cite{Kibble:1976,Hindmarsh:1995}.  In the familiar four-dimensional case ($n=4$) one typically chooses
  $\eta \sim10^{16}\,\mathrm{GeV}\simeq M_{P}
  =\ell_{P}^{-1}$, with $M_{P}$ and $\ell_{P}$ the Planck mass and length, respectively\,\cite{Burgess:2007}.  More generally, dimensional analysis in $n$ spacetime dimensions implies that $\eta$ carries mass dimension one but is bounded by the natural short-distance cutoff set by the $n$-dimensional Planck scale: $\eta \;\lesssim\; \ell_{P}^{-(n-2)/2}$.

By varying the action~\eqref{action} with respect to the metric tensor, we derive the corresponding FEs:
\small
 \begin{eqnarray}
 \mathcal{I}_{\mu\nu}=\mathcal{R}_{\mu\nu}-\frac{1}{2}g_{\mu\nu}(\mathcal{R}-2\Lambda)-\kappa\left(\mathcal{T}_{\mu\nu}^{\text{YM}}+\mathcal{T}_{\mu\nu}^{\text{GM}}+\mathcal{T}_{\mu\nu}^{\text{Mur}}\right)&=&0,\hspace{0.6cm}\label{eom}
 \end{eqnarray}
 \normalsize
where $\mathcal{T}_{\mu\nu}^{\rm YM}$, $\mathcal{T}_{\mu\nu}^{\rm GB}$, and $\mathcal{T}_{\mu\nu}^{\rm Mur}$ denote, respectively, the energy-momentum (em) tensors associated with the PYM field, GM, and matter with a Murnaghan-type EoS. To make significant progress in solving the appropriate FEs, we first focus on the PYM sector, where $\mathcal{T}_{\mu\nu}^{\rm YM}$ can be expressed explicitly as:
\begin{equation}
    \mathcal{T}_{\mu}^{\nu\,\rm YM}=-\frac{1}{2}\Bigg(\delta_\mu^\nu\mathcal{F}^\epsilon-4\,\epsilon\,\sum_{i=1}^{(n-2)(n-1)/2} \left(F_{\mu\lambda}^{(i) }F^{(i)\nu\lambda}\right)\mathcal{F}^{\epsilon-1}\Bigg).
\end{equation}
In a similar vein, the variation of the action~\eqref{action} with respect to the gauge potentials $A^{(i)}$ produces the following result.
\begin{equation}
    \mathrm{d}\left(^\star F^{(i)}\mathcal{F}^{\epsilon-1}\right)+\frac{1}{\sigma}C^{(i)}_{(j)(k)}\mathcal{F}^{\epsilon-1}A^{(j)}\wedge\,^\star F^{(k)}=0 \,,
\end{equation}
where $\star$ represents the duality.
Similarly, the em tensor associated with the GM configuration, which arises from the variation of the underlying action functional with respect to $g^{\mu\nu}$, is supplemented by the dynamical equation of the SF expressed in the following explicit form:
\begin{eqnarray}
\mathcal{T}_\mu^{\nu\,\rm GM}&=&\partial_\mu\phi^a\partial^\nu\phi^a-\delta_\mu^\nu\left(\frac{1}{2}(\partial\phi^a)^2+\mathcal{V}(\phi^2)\right),\\
\square\phi^a&=&\frac{\partial \mathcal{V}}{\partial\phi^a}.
\end{eqnarray}

To obtain an exact BH solution in a $n$-dimensional gravitational theory, one begins by specifying a metric ansatz that is manifestly adapted to the symmetries of the problem. Focusing on a static, spherically symmetric spacetime, we impose the simplifying gauge $g_{tt}\,g_{rr} = -1$, which collapses the independent metric degrees of freedom to a single radial lapse. Consequently, the line element may be written in the canonical one-function form
\begin{align}
    \mathrm{d}s^2 = -F(r)\,\mathrm{d}t^2 + \frac{\mathrm{d}r^2}{F(r)} + r^2\,\mathrm{d}\Omega_{n-2}^2, \label{met}
\end{align}
where $F(r)$ is an undetermined metric function encoding the gravitational dynamics, and $\mathrm{d}\Omega_{n-2}^2$ denotes the line element of a hypersurface of a maximally symmetric $(n-2)$ dimensional $\mathrm{d}\Omega_{n-2}^2$ of constant curvature $(n-2)(n-3)$.

The hypersurface $\mathrm{d}\Omega_{n-2}^2$, corresponding to the angular part of the geometry, can be explicitly parametrized in spherical coordinates as
\begin{align}
    \mathrm{d}\Omega_{n-2}^2 = \mathrm{d}\theta^2 + \sin^2\theta \bigg[ \mathrm{d}\phi_1 + \sum_{i = 2}^{n-3} \prod_{j = 1}^{i-1} \sin^2\phi_j\, \mathrm{d}\phi_i^2 \bigg],
\end{align}
with the angular coordinate $ \theta \in [0, \pi/2] $, and the remaining angles $ \phi_i $ defined over their standard ranges. This structure ensures that the angular sector of the spacetime corresponds to a $(n-2)$-dimensional unit sphere.

For convenience in expressing quantities in Cartesian-like coordinates adapted to the higher-dimensional sphere, one may define the embedding coordinates $ \{x_i\} $ as
\begin{align}
    x_1 &= r \cos\theta, \\
    x_i &= r \sin\theta \cos\phi_{n - i - 1} \prod_{j=1}^{n - i - 2} \sin\phi_j, \quad i = 2, \ldots, n-2, \\
    x_{n-1} &= r \sin\theta \prod_{j=1}^{n - 4} \sin\phi_j,
\end{align}
where $ r $ denotes the areal radius of the hypersphere. These coordinate definitions are instrumental in embedding the angular part of the metric into a higher-dimensional Euclidean space, which becomes particularly relevant in the analysis of topological properties, global charges, and horizon structures of the BH solution.

We consider a static, spherically symmetric spacetime minimally coupled to an anisotropic fluid whose em tensor is given by \cite{Raposo:2018rjn}
\begin{equation}
    \mathcal{T}_{\mu\nu} = (\rho + p_t) u_\mu u_\nu + p_t g_{\mu\nu} + (p_r - p_t) \chi_\mu \chi_\nu\,,
    \label{eq:Tmunu}
\end{equation}
where $\rho$ is the energy density, $p_r$ and $p_t$ are the radial and tangential pressures, respectively, $u^\mu$ is the $n$-velocity of the fluid and $\chi^\mu$ is a unit space-like vector orthogonal to $u^\mu$. These vectors satisfy the normalization conditions $u^\mu u_\mu = -1$, $\chi^\mu \chi_\mu = 1$, and $u^\mu \chi_\mu = 0$.

This general anisotropic structure extends the usual perfect-fluid description and is capable of modeling DE configurations with directional pressure disparities due to strong curvature effects. Anisotropies are known to emerge naturally in a wide range of contexts, including in the presence of magnetic fields, phase transitions, or quantum fields in curved backgrounds \cite{Herrera:1997plx,Bowers:1974tgi,Chan:1993mn}.

In the comoving frame with $u^\mu = \sqrt{F(r)}\, \delta_0^\mu$ and $\chi^\mu = 1/\sqrt{F(r)}\, \delta_1^\mu$, the mixed components of the em tensor reduce to
\begin{equation}
    \mathcal{T}^\mu_{\;\nu} = \text{diag}(-\rho, p_r, p_t, \dots, p_t)\,,
    \label{eq:Tmunu_diag}
\end{equation}
clearly displaying the anisotropic nature of the stress distribution.

Rather than adopting the standard barotropic relation $p = \omega \rho$, we employ a non-linear EoS inspired by the Murnaghan form \cite{Murnaghan:1944}: 
\begin{equation}
    p(\rho) = -\beta \rho - \gamma \rho^{1 + \frac{1}{\xi}}\,,
    \label{Murnaghan}
\end{equation}
where the constants $\beta > 0$, $\gamma > 0$, and $\xi > 0$ are model parameters. This EoS mimics a DE fluid with a pressure more negative than that of a cosmological constant, corresponding to a phantom-like regime \cite{Caldwell:2002,Nojiri:2005}.

In fact, the form \eqref{Murnaghan} generalizes the linear phantom fluid model ($p = \omega \rho$ with $\omega < -1$) and introduces non-linear corrections relevant in high-density regimes, possibly arising from QG effects or exotic matter configurations. It is also consistent with recent effective field theory descriptions of DE fluids in strongly curved spacetimes \cite{Babichev:2004yx,Bertacca:2010ct}.

In Schwarzschild-like spacetimes, the radial coordinate becomes temporal inside the event horizon. To maintain regularity and continuity across the horizon, we impose the constraint:
\begin{equation}
    p_r = -\rho\,,
    \label{eq:prho}
\end{equation}
which guarantees that the energy density remains invariant under $r \leftrightarrow t$ coordinate exchange. This condition has been employed in several models of anisotropic DE accretion and BH interiors \cite{Kiselev:2002dx,Babichev:2004yx}.

To relate the tangential pressure $p_t$ to the averaged isotropic pressure $p$ obeying the Murnaghan EoS, we use the angular averaging condition:
\begin{equation}
    p = p_t + \frac{1}{n - 1}(p_r - p_t)\,,
    \label{eq:avg_p}
\end{equation}
which arises from projecting the em tensor on an angular hypersurface: $\langle \delta_i^1 \delta_1^j \rangle = 1/(n - 1)$. Substituting Eq.~\eqref{eq:prho} and the EoS \eqref{Murnaghan} into Eq.~\eqref{eq:avg_p}, one finds that:
\begin{equation}
    p_t = \frac{1}{n - 2} \left( (1 - (n - 1)\beta)\rho - (n - 1)\gamma \rho^{1 + \frac{1}{\xi}} \right)\,.
    \label{eq:pt}
\end{equation}

The final expressions for the nonzero mixed components of the em tensor are:
\begin{align}
    \mathcal{T}^t_{\;t} &= \mathcal{T}^r_{\;r} = -\rho\,, \label{eq:Ttt} \\
    \mathcal{T}^{\theta_i}_{\;\theta_i} &= p_t = \frac{1}{n - 2} \left( (1 - (n - 1)\beta)\rho - (n - 1)\gamma\, \rho^{1 + \frac{1}{\xi}} \right)\,.
    \label{eq:Ttheta}
\end{align}
These expressions ensure that the effective fluid exhibits repulsive gravitational behavior and anisotropy consistent with a phantom DE model. The degree of anisotropy is determined by the deviation of $p_t$ from $p_r = -\rho$ and is controlled by the parameters $\beta$, $\gamma$, and $\xi$. The condition $p_t > p_r$ can lead to dynamical instabilities or exotic horizon structures, while $p_t < p_r$ may support stable compact configurations depending on boundary conditions \cite{Chan:1993mn, Herrera:1997plx}.

\section{Energy density profile and spacetime solution}
\label{sec2}
The study of a precise BH solution requires looking at the equations that describe gravity and matter \eqref{eom}. By considering static and spherically symmetric features \eqref{met}, one finds that $\mathcal{T}_t^t = \mathcal{T}_r^r$. Consequently, the gravitational framework can be modeled in higher-dimensional spacetime through the following FEs:
    \begin{eqnarray}
\mathcal{I}_t^t=\mathcal{I}_r^r&=&\frac{1}{2r^2}(n-2)(n-3)(F(r)-1)+\frac{1}{2r}(n-2)F'(r)+\Lambda,\label{g1}\\
     \mathcal{I}_{\theta_i}^{\theta_i}= \mathcal{I}_{\theta_1}^{\theta_1}&=&\frac{F''(r)}{2}+\frac{(F(r)-1)(n-3)(n-4)}{2r^2}+\frac{(n-3)F'(r)}{r}+\Lambda.\label{g2}
    \end{eqnarray}
    
At this juncture, it may be beneficial to adopt a more thorough approach by examining the matter sector associated with non-linear electrodynamics, GMs, and the matter fluid using the Murnaghan EoS. In this way, by using the line element \eqref{met} and the WY ansatz from Refs. \cite{WuYang:1969,Naeimipour:2021bgc,Mazharimousavi:2009,Ali:2019lzh}, the PYM FEs \cite{HabibMazharimousavi:2008dm,Hassaine:2007py} are satisfied if the gauge potential one-forms are expressed as
\begin{equation}
    A^{(i)}=\frac{Q}{r^2}C_{(l)(j)}^{(a)}x^l\mathrm{d}x^j,\,\, r^2=\sum_{l=1}^{n-1}x_l^2 \,,
\end{equation}
where $Q$ matches the YM magnetic charge \cite{WuYang:1969,Yang:1974}, and the index range satisfies the hierarchy $2 \leq j + 1 \leq l \leq n - 1$. Under this ansatz, one finds
\begin{equation}
\mathcal{F}
=\frac{(n-2)(n-3)\,Q^2}{r^4}
\,,\qquad
\sum_{i=1}^{(n-2)(n-1)/2} \left(F_{\mu\lambda}^{(i) }F^{(i)\mu\lambda}\right)
=\frac{(n-3)\,Q^2}{r^4}
=\frac{1}{n-2}\,\mathcal{F}\,,
\end{equation}
so that the YM em tensor diagonalizes in an orthonormal frame to
\begin{equation}
T^i{}_{j}
=-\tfrac12\,\mathcal{F}^\epsilon\,\mathrm{diag}\bigl[1,\,1,\,\alpha,\,\dots,\alpha\bigr]
\,,\quad
\alpha=1-\tfrac{4\epsilon}{n-2}
\,,\end{equation}
which reveals an explicit anisotropy between the radial and transverse pressures \cite{BowersLiang:1974,Herrera:1997plx}. Its trace is 
$T=T^\mu{}_{\mu}=-(1/2)\,\mathcal{F}^\epsilon\,(N-4\epsilon)$,
which vanishes precisely when $\epsilon=N/4$, corresponding to a conformal (traceless) YM source and underpinning a class of extremal and superconformal BH solutions in AdS/CFT and higher-dimensional supergravity contexts \cite{CallanColemanJackiw,AwadJohnson2000}.

GM configurations within GR can be treated in terms of the Higgs profile $\gamma(r)$ and the metric spacetime elements, which embrace no exact exterior solution and ought to be integrated numerically.  Nonetheless, because the SF eventually approaches its vacuum expectation value (VEV) outside a core of radius $\delta \sim \eta^{-2/(n-2)} \ll r_h$. In addition, the thin-core approximation may be used when the energy density in the outer area scales as $\rho\sim\eta^2/r^2$ \cite{BV1989,BarrosRomero1997}. For $r\ge\delta$ (and thus outside the horizon at $r_h$), one set
$\gamma(r)=1$ and $\mathcal{V}\bigl(\phi^a\bigr)=0$, so that the stress-energy tensor reduces to
\begin{equation}
    T_\mu^{\nu\,GM}=\left\lbrace T_0^0,T_1^1,T_i^i\right\rbrace=\left\lbrace -\frac{(n-2)}{2r^2}\eta^2,-\frac{(n-2)}{2r^2}\eta^2,-\frac{(n-4)}{2r^2}\eta^2\right\rbrace \,,
\end{equation}
yielding a spacetime with a solid-angle deficit but no Newtonian gravitational attraction \cite{BV1989,Preskill1993}.  The numerical interior solution for $r<\delta$ is then matched at $r\approx\delta$ to this analytic exterior form, ensuring continuity of the metric and its first derivatives across the core boundary \cite{Rahaman2009,Zhang2025}.

In seeking an exact, analytical solution in GR, one must simultaneously satisfy the gravitational and the matter FEs \eqref{eom}.  Equivalently, the contracted Bianchi identity such that $T^{\mu\nu}{}_{;\nu}=0$, yields the fluid’s continuity equation in a static, spherically symmetric background \cite{Tolman1939,Oppenheimer1939}.  Upon substituting the Murnaghan‐type EoS \eqref{Murnaghan}, one obtains directly the first-order radial ordinary differential equation
\begin{equation}\label{murn}
   r\rho'(r) +(n-1)\rho(r)-(n-1)\bigg(\beta\,\rho(r)+\gamma\,\rho(r)^{1+\tfrac{1}{\xi}}\bigg)=0\,. 
\end{equation}
 \begin{figure*}[tbh!]
      	\centering{
       \includegraphics[height=6.5cm,width=7.5cm]{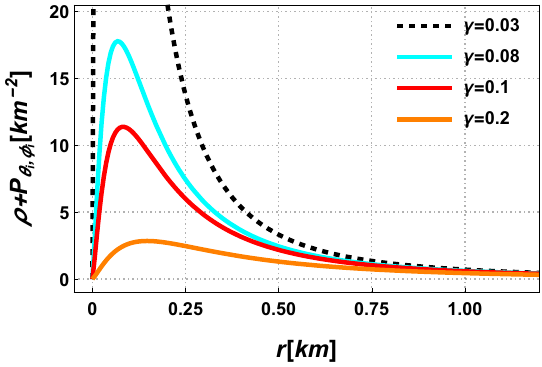} \hspace{1cm}
      	\includegraphics[height=6.5cm,width=7.8cm]{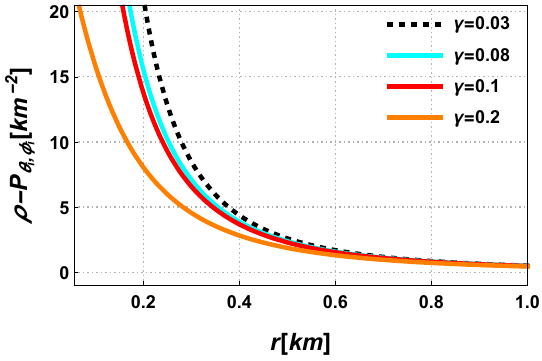} \hspace{1cm}
       \includegraphics[height=6.5cm,width=8cm]{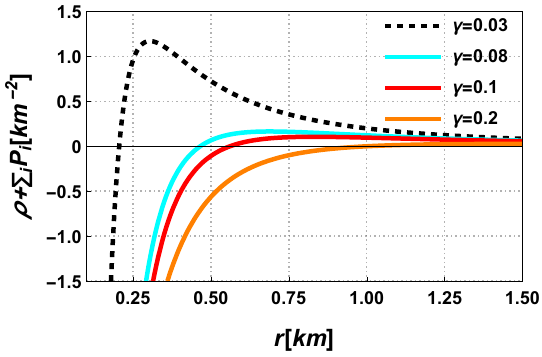}
      }
      	\caption{The variation of $\rho+\sum_i p_i$ (SEC), $\rho+p_{\theta_i,\phi_i} $ (NEC), and $\rho-\mid p_{\theta_i,\phi_i}\mid $ (DEC) against $r[Km]$ for multiple values of the Murnaghan modulus $\gamma$ and for the fixed set, i.e., $\beta=0.2$, $\xi=2$, $\chi=1$, and $n=4$.}\label{ECs1}
      \end{figure*}
       \begin{figure*}[tbh!]
      	\centering{
       \includegraphics[height=6.5cm,width=7.5cm]{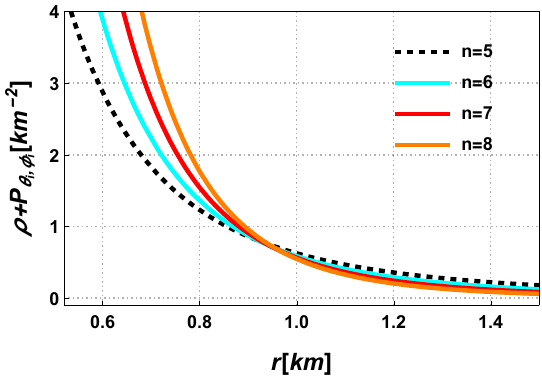} \hspace{1cm}
      	\includegraphics[height=6.5cm,width=7.8cm]{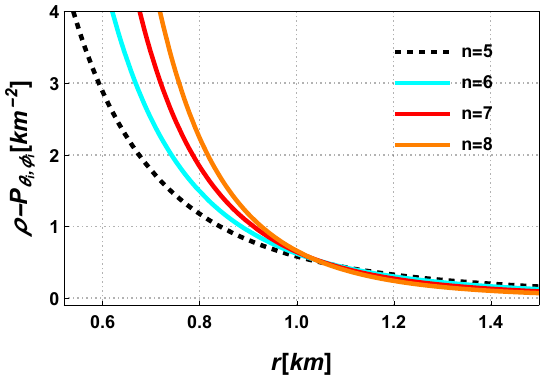} \hspace{1cm}
       \includegraphics[height=6.5cm,width=8cm]{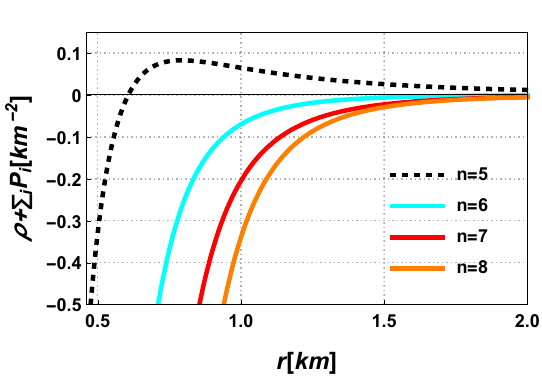}
      }
      	\caption{The variation of $\rho+\sum_i p_i$ (SEC), $\rho+p_{\theta_i,\phi_i} $ (NEC), and $\rho-\mid p_{\theta_i,\phi_i}\mid $ (DEC) against $r[Km]$ for multiple spacetime dimensions and for the fixed set, i.e., $\gamma=0.03$, $\beta=0.2$, $\chi=1$, and $\xi=2$.}\label{ECs2}
      \end{figure*}
Here the term $r\,\rho'(r)$ encodes the geometric dilution in $n+1$ spatial dimensions, while $(n-1)\,\beta\rho$ and $(n-1)\gamma\,\rho^{1+1/\xi}$ respectively quantify the linear and non-linear pressure back reaction of the Murnaghan fluid \cite{Murnaghan:1944,Wallace:1972}.  Eq. ~\eqref{murn} thus generalizes the familiar polytropic density profile equation by incorporating a density-dependent bulk modulus, and it governs the fall-off of $\rho(r)$ outside the monopole core, interpolating between Hookean and strongly non-linear regimes.  By integrating Eq.~\eqref{murn} using the Murnaghan EoS, one obtains a closed-form expression for the energy density profile of the generalized polytropic fluid, such that
\begin{equation}
\rho(r) =\left(\frac{\gamma +\chi ^2 r^{-\frac{(\beta -1) (n-1)}{\xi}}}{1-\beta }\right)^{-\xi },\label{ho}
\end{equation}
where $\chi$ serves as an integration constant in the continuity equation, representing the amplitude of the non-linear density contribution within $\rho(r)$. A closer examination shows that for $r\ll r_c$, where $r_c=\bigl[(1-\beta)\chi^2/\gamma\bigr]^{1/\zeta}$ with $\zeta = (n-1)(\beta-1)/\xi$, the term $\chi^2r^{-\zeta}\gg\gamma$ dominates, yielding a steep core profile $\rho\propto r^{(n-1)(\beta-1)}$ characteristic of higher-dimensional anisotropic fluids \cite{Tolman1939,Oppenheimer1939,Herrera:1997plx}.  As $r\to\infty$, the constant $\gamma$ term prevails and $\rho\to[\gamma/(1-\beta)]^{-\xi}$, an effective vacuum energy driving late-time acceleration \cite{Caldwell:2002,Nojiri:2005}.  Positivity and regularity require $\gamma>0$ and $1-\beta>0$, while the phantom regime $\beta>1$ (so $w=p/\rho<-1$) entails a mild null EC violation consistent with BH phantom accretion models \cite{Babichev:2004yx}.  Hence, this analytic form unifies non-linear core dynamics with a DE‐like exterior in a single Murnaghan-type profile.

 Alternatively, selecting $r\rightarrow \infty$ leads to the asymptotic forms of the pressure components becoming
\begin{align}
p_{r,\infty} &= -\beta\,\rho_\infty - \gamma\,\rho_\infty^{1 + \tfrac{1}{\xi}}, \\
p_{t,\infty} &= \tfrac{1}{n - 2} \left((n - 1)\,p_\infty - p_{r,\infty}\right) 
= \tfrac{1}{n - 2} \left((n - 2)(-\beta\,\rho_\infty) - \gamma\,\rho_\infty^{1 + \tfrac{1}{\xi}}\right),
\end{align}
where $p_\infty = -\beta\,\rho_\infty - \gamma\,\rho_\infty^{1 + \tfrac{1}{\xi}}$. This leads to a nonvanishing anisotropy factor,
$
\Delta_\infty = p_{t,\infty} - p_{r,\infty} 
\neq0$, which only vanishes if
$\gamma = (1 - \beta)\,\rho_\infty^{-1/\xi}$. 
This behavior contrasts with quintessential and Chaplygin dark fluid (CDF) models, in which $\Delta \to 0$ at large distances \cite{Kiselev:2002dx, Sekhmani:2024udl}. The persistence of anisotropy in the Murnaghan fluid may yield observable consequences for large-scale structure and cosmic expansion, such as anisotropies in the Hubble flow and galaxy cluster dynamics \cite{Migkas:2017vir, Migkas:2020fza, DelCampo:2012wkb}, thereby warranting deeper investigation into the role of Murnaghan-type DE in gravitational systems.

To gain clearer insight into the behavior of our solution, we now examine the classical ECs, namely the null EC (NEC), dominant EC (DEC), weak EC (WEC), and strong EC (SEC), which are defined as follows~\cite{Kontou:2020bta}:
\begin{eqnarray}\label{18}
\textbf{WEC} &:& \rho \geq 0,\quad \rho + P_i \geq 0, \,\,
\textbf{SEC} : \rho + \sum_i P_i \geq 0,\quad \rho + P_i \geq 0, \nonumber\\[0.5cm]
\textbf{NEC} &:& \rho + P_i \geq 0, \,\,
\label{19}
\textbf{DEC} : \rho \geq 0,\quad |P_i| \leq \rho.
\label{20}
\end{eqnarray}
Thus, complete expressions may be provided as follows:
\begin{align}
   \rho + p_r =0,\,\, \rho+ P_{\theta_{i},\phi_{i}}&=\frac{n-1}{n-2}\left(\rho\,(1-\beta)-\gamma\,\rho^{1+\frac{1}{\xi}}\right),\nonumber\\
   \rho+ \sum_{i}p_i&=\rho\,\left(1-\beta\left(\frac{n-1}{n-2}\right)\right)-\left(\frac{n-1}{n-2}\right)\,\gamma\,\rho^{1+\frac{1}{\xi}}\,,\\
   \rho-|p_r|=0,\,\,  \rho-|p_{\theta_{i},\phi_{i}}|&=\rho-\Bigl|\frac{\rho}{n-2}\left(\frac{1-\beta(n-1)}{n-2}\right)-\left(\frac{n-1}{n-2}\right)\gamma\,\rho^{1+\frac{1}{\xi}}\Bigr|\,.\nonumber
\end{align}
Investigating the ECs according to the exact solution for the energy density \eqref{ho} yields the following explicit expressions:
    \begin{align}
\rho+ P_{\theta_{i},\phi_{i}}&=-\frac{(\beta -1) (n-1) \chi ^2 \left(\frac{\gamma +\chi ^2 r^{-\frac{(\beta -1) (n-1)}{\xi }}}{1-\beta }\right)^{-\xi }}{(n-2) \left(\gamma 
   r^{\frac{(\beta -1) (n-1)}{\xi }}+\chi ^2\right)},\nonumber\\
   \rho+ \sum_{i}P_i&=\left(\frac{\gamma +\chi ^2 r^{-\frac{(\beta -1) (n-1)}{\xi }}}{1-\beta }\right)^{-\xi } \left(\beta(1 -n)-\frac{(1-\beta ) \gamma 
   (n-1)}{\gamma +\chi ^2 r^{-\frac{(\beta -1) (n-1)}{\xi }}}+1\right),\\
\rho-|P_{\theta_{i},\phi_{i}}|&=\frac{\left(\frac{\gamma +\chi ^2 r^{-\frac{(\beta -1) (n-1)}{\xi }}}{1-\beta }\right)^{-\xi } \left(-\beta(1 + n)+\frac{(1-\beta ) \gamma 
   (n-1)}{\gamma +\chi ^2 r^{-\frac{(\beta -1) (n-1)}{\xi }}}+n-3\right)}{n-2}\,.\nonumber
\end{align}

In the framework of our exact solution with $\xi = 2$, the classical ECs provide significant insights into the nature of the matter sector. The NEC, expressed as $\rho + P_{\theta_i,\phi_i}$, is satisfied when $\beta < 1$, marginally satisfied for $\beta = 1$, and violated when $\beta > 1$. Such violations typically suggest the presence of exotic matter fields, such as those required to support traversable wormholes~\cite{Morris:1988cz,Sushkov:2005kj}. The SEC, given by $\rho + \sum_i P_i$, exhibits a more sensitive dependence on both $\beta$ and the spacetime dimension $n$, and is generally violated for $\beta > 1/n-1$. This is consistent with scenarios involving late-time cosmic acceleration or inflation, where SEC violation is expected~\cite{Barcelo:1999hq}. The DEC, formulated as $\rho - |P_{\theta_i,\phi_i}|$, is satisfied near the origin provided $\beta < (n - 3)/(n + 1)$. At spatial infinity, the constraint becomes less stringent; for instance, when $n = 4$, the condition is upheld for $\beta \lesssim 0.7$. Violation of the DEC is typically interpreted as allowing negative energy densities or superluminal energy fluxes~\cite{Hawking:1973uf}. Notably, in the regime $\beta > 1$, all ECs are simultaneously violated, indicating that the matter behaves as an effective phantom or non-canonical field, in line with several models of modified gravity or anisotropic compact stars~\cite{Lobo:2005vc,Nojiri:2010wj}. In contrast, for $\beta \in (0, 0.3)$, particularly when $\gamma$ is small, all ECs are satisfied over a wide range of $r$, suggesting physically viable matter configurations. Broadly speaking, these violations are a generic feature of viable DE models in the late-time universe or near BH horizons \cite{Carroll:2003st}. From a graphical perspective, Figures \ref{ECs1}-\ref{ECs2} illustrate the variation of the NEC, DEC, and SEC in relation to the radial coordinate $r$ across different values of the Murnaghan parameter space $(\beta,\gamma,\chi,n)$. In Figure \ref{ECs1} (varying $\gamma$ for $n=4$), the tangential NEC, $\rho + P_{\theta_i,\phi_i}$, remains positive throughout the exterior region for all considered $\gamma$, indicating that the NEC is well satisfied~\cite{Barcelo:1999hq,Visser:2000at}. Similarly, the DEC indicator, $\rho - |P_{\theta_i,\phi_i}|$, stays positive beyond the near-horizon region for small to moderate $\gamma$, confirming causal energy flow and the absence of superluminal or negative-energy stresses in most of the spacetime~\cite{Hawking:1973uf,Lobo:2005vc}; minor dips near the core for larger $\gamma$ are limited and do not undermine the overall satisfaction of DEC. Conversely, $\rho + \sum_i P_i$ remains negative for all $\gamma$, confirming universal SEC violation and thus repulsive gravitational behavior at both local and cosmological scales~\cite{Nojiri:2010wj,Caldwell:2002}. In Figure \ref{ECs2} (fixed $\gamma=0.008$, varying $n$), increasing the spacetime dimension from 5 to 8 gradually raises the tangential NEC curve, further reinforcing NEC satisfaction over a wider radial range~\cite{Migkas:2020fza,Migkas:2017vir}. The DEC curve’s zero crossing shifts inward with increasing $n$, illustrating improved restoration of causality bounds in higher dimensions~\cite{Germani:2001du,Maartens:2003tw}. Although the SEC combination also becomes less negative with dimension, it never becomes positive, indicating persistent but dimensionally weakened repulsive effects. These results collectively demonstrate that while the Murnaghan anisotropic fluid consistently satisfies NEC and DEC across most of the exterior spacetime, it robustly violates SEC, reflecting its phantom-like and repulsive gravitational character~\cite{Sushkov:2005kj,Lobo:2005vc}; moreover, increasing spacetime dimension systematically mitigates-though does not completely remove the exotic aspects of the model.
 
   	At this stage, with all the exact solutions for the matter sector available, the process of modeling the exact analytical BH solution is examined by taking into account the following complete set of FEs
 \small
    \begin{eqnarray}
\mathcal{I}_t^t=\mathcal{I}_r^r&=&\frac{1}{2r^2}(n-2)(n-3)(F(r)-1)+\frac{(n-2)}{2r}F'(r)+\Lambda+\frac{1}{2}\left(\frac{(n-2)(n-3)Q^2}{r^4}\right)^\epsilon+\frac{(n-2)}{2r^2}\eta^4+\rho(r),\hspace{0.75cm}\label{g1a}\\
     \mathcal{I}_{\theta_i}^{\theta_i}= \mathcal{I}_{\theta_1}^{\theta_1}&=&\frac{F''(r)}{2}+\frac{(F(r)-1)(n-3)(n-4)}{2r^2} +\frac{(n-3)F'(r)}{r} +\Lambda+\frac{1}{2}\left(\frac{(n-2)(n-3)Q^2}{r^4}\right)^\epsilon\left(1-\frac{4\epsilon}{n-2}\right)\nonumber\\
     &+&\frac{(n-4)}{2r^2}\eta^2+\rho(r)\left(\frac{1-\beta(n-1)}{n-2}\right)+\frac{n-1}{n-2}\gamma\,\rho^{1+\frac{1}{\xi}} .\label{g2a}
    \end{eqnarray}
\normalsize
By considering the energy density according to the Murnaghan EoS \eqref{ho}, along with the $\small\bigg( \begin{array}{c}t \\t\end{array}\small\bigg)$ components of the FEs, one can derive the following ordinary differential equation:
\begin{figure*}[tbh!]
      	\centering{
       \includegraphics[height=7cm,width=8.0cm]{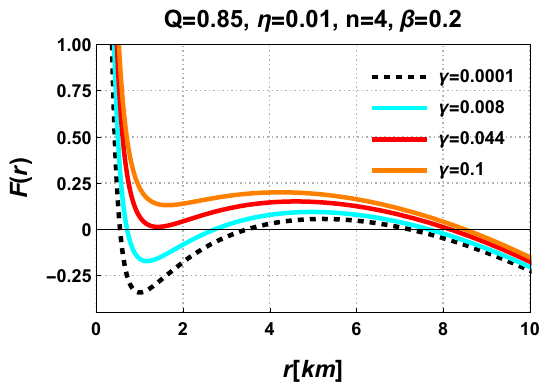} \hspace{1cm}
      	\includegraphics[height=7cm,width=8.0cm]{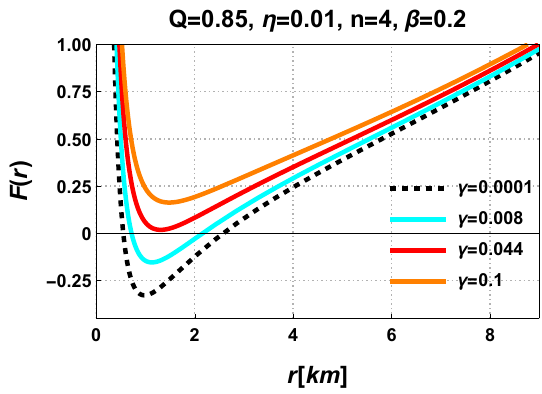} \hspace{1cm}
       \includegraphics[height=7cm,width=8.0cm]{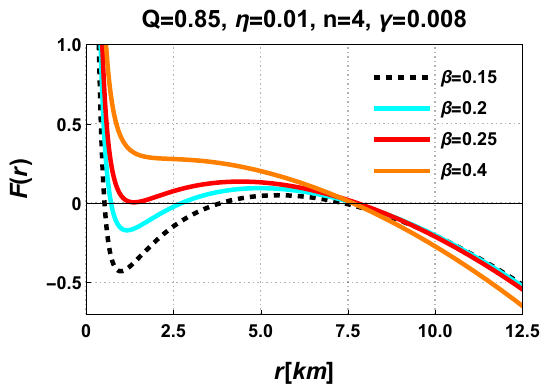}
       \hspace{1cm}
       \includegraphics[height=7cm,width=8.0cm]{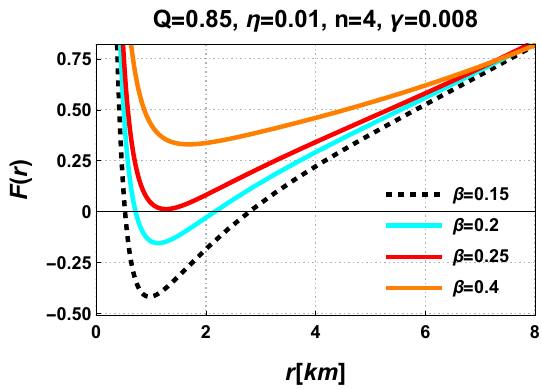}
      }
       
      	\caption{The metric function $F(r)$ plots versus {$r[km]$} for various values of the Murnaghan modulus $\eta$ and the linear (barotropic) coefficient $\beta$. The right panel corresponds to the AdS configuration with $(\Lambda=-0.02,\,\chi=1)$, and the left panel to the dS configuration with $(\Lambda=0.02,\,\chi=1)$.} 
    \label{fig3a}
      \end{figure*}

\begin{figure*}[tbh!]
      	\centering{
       \includegraphics[height=7cm,width=8.0cm]{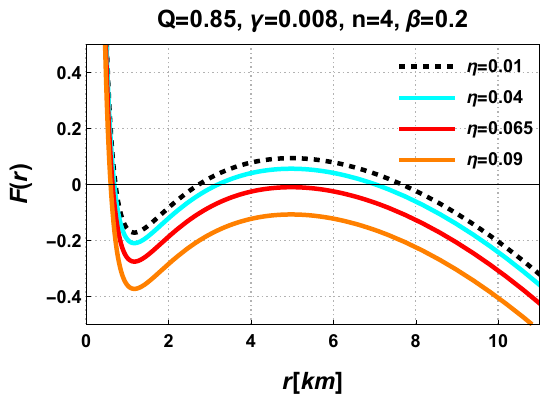} \hspace{1cm}
      	\includegraphics[height=7cm,width=8.0cm]{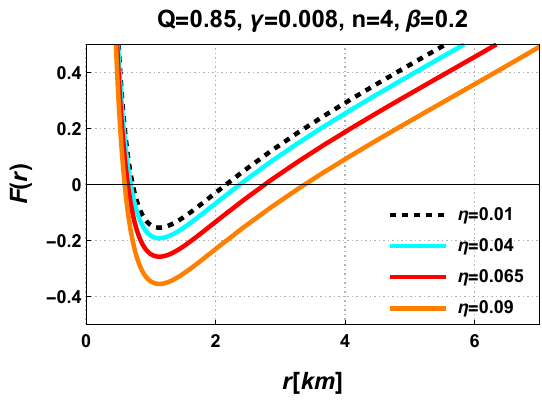} \hspace{1cm}
       \includegraphics[height=7cm,width=8.0cm]{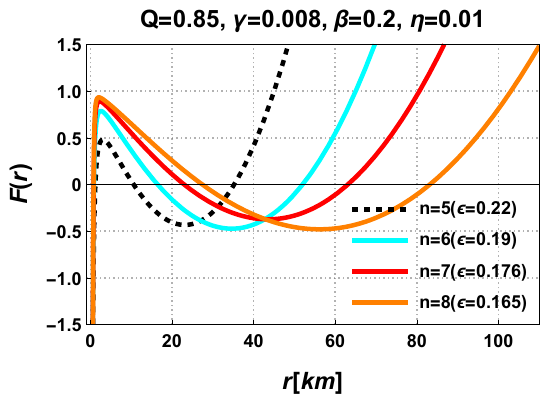}
       \hspace{1cm}
       \includegraphics[height=7cm,width=8.0cm]{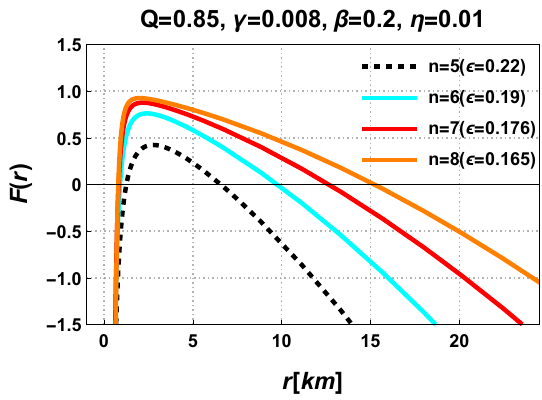}
      }
       
      	\caption{The metric function $F(r)$ plots versus {$r[km]$} for various values of the GM charge $\eta$ and the spacetime dimensions $n$. The right panel corresponds to the AdS configuration with $(\Lambda=-0.02,\,\chi=1)$, and the left panel to the dS configuration with $(\Lambda=0.02,\, \chi=1)$.} 
    \label{fig4a}
      \end{figure*}

     \begin{figure*}[tbh!]
      	\centering{
       \includegraphics[scale=0.57]{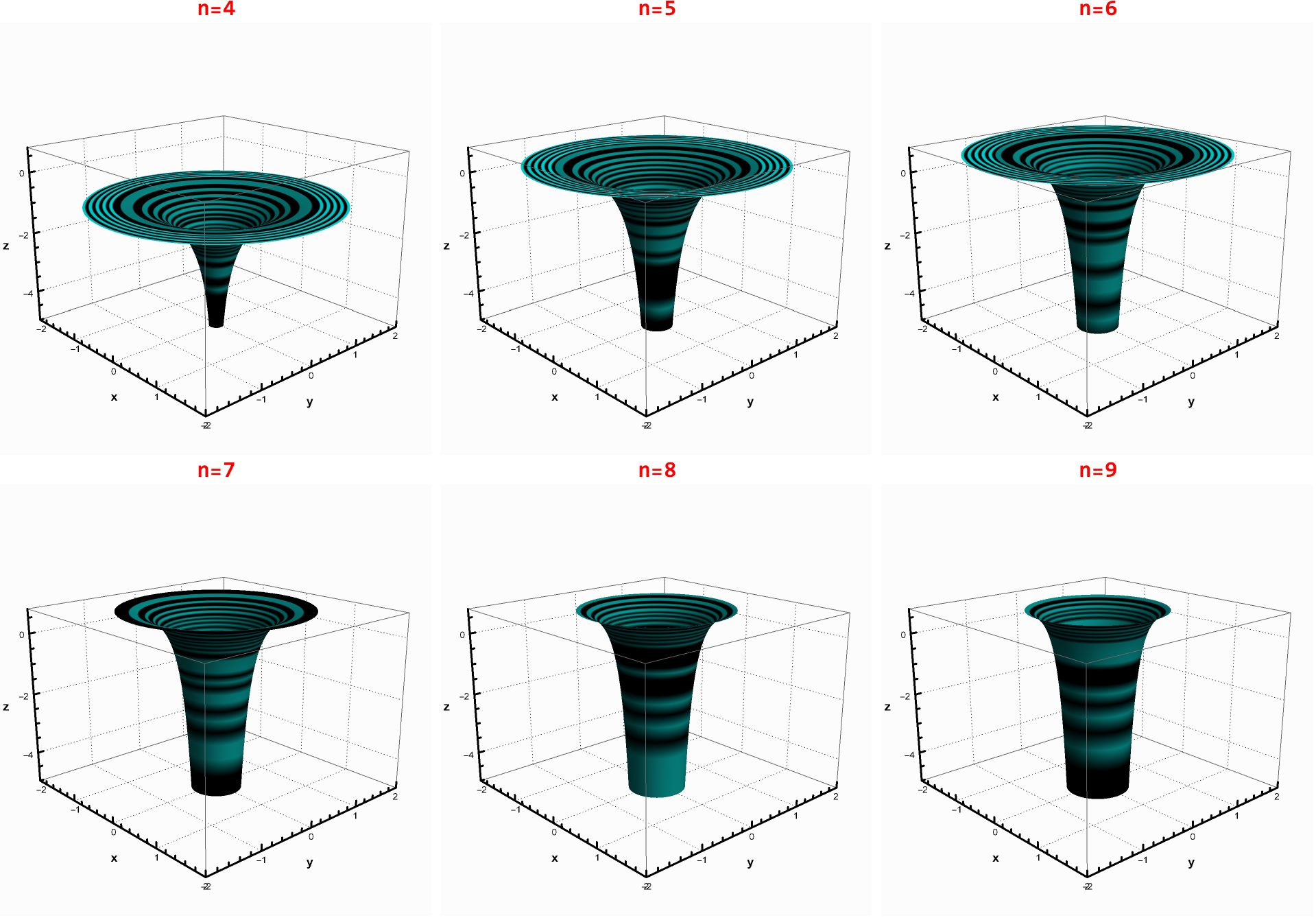} \hspace{1cm}
      }
       
      	\caption{The profile of the embedding diagram within different spacetime dimensions $n$, where  $\Lambda = -0.001,  
Q = 0.8, 
\epsilon = 0.1, 
\beta = 0.09, 
\gamma = 0.009, 
\xi=2,
\chi = 1, 
\eta = 0.01.$, and the BH mass has been set to unit $M=1$} 
    \label{emm}
      \end{figure*}

    \begin{align}
   \bigg( \frac{n-2}{2 r^2}\bigg)\bigg(r F'(r)+(n-3)( F(r)-1)\bigg)+\Lambda +\frac{1}{2} \left(\frac{(n+1) (n+2) Q^2}{r^4}\right)^{\epsilon
   }+\frac{(n-2)}{2r^2}\eta^4+\left(\frac{\gamma +\chi ^2 r^{-\frac{(\beta -1) (n-1)}{\xi }}}{1-\beta }\right)^{-\xi }=0 ,
    \end{align}
in which an exact analytical expression for $F(r)$ is given by the following form
   \small
    \begin{align}\label{metric}
 \hspace{-0.5cm} F(r) =1-\frac{16 \pi  M r^{2-n}}{(n-1) \Omega _{n-1}}-\frac{2 \Lambda  r^2}{n^2-3 n+2}-\frac{r^2  \left(\frac{32 \pi ^{2  }Q^2}{r^4 \Omega _{n-1}^2}\right)^{\epsilon }}{(n-1) (n-4
   \epsilon )}-\frac{8
   \pi  \eta ^2}{n-3}-\frac{2 r^2 \left(\frac{1-\beta }{\gamma }\right)^{\xi } \, _2F_1\left(\xi ,\frac{\xi }{1-\beta };\frac{\xi }{1-\beta
   }+1;-\frac{r^{-\frac{n (\beta -1)}{\xi }} \chi ^2}{\gamma }\right)}{n(n-1)} \,,
\end{align}
\normalsize
where $\epsilon\neq\frac{n}{4}$. Here, $ M $ and $ Q $ denote, respectively, the ADM mass \cite{Ashtekar:1984zz,Ashtekar:1999jx} and the total YM magnetic charge, given by
\begin{equation}
M = \frac{(n - 2)\, \omega_{n-2}}{16\pi}\, m, \qquad
Q = \frac{1}{4\pi} \int d^{n-2}x\, \sqrt{\sum_{i=1}^{(n-2)(n-1)/2} \left(F_{\mu\lambda}^{(i) }F^{(i)\mu\lambda}\right)}
= \sqrt{\frac{(n - 2)(n - 3)\, \omega_{n-2}}{4\pi}}\, e,
\end{equation}
where $m$ is an integration constant and $\omega_{n-2}$ is the volume of the unit $(n-2)$-dimensional sphere. Moreover, the Gaussian hypergeometric function ${}_2F_1(a_1,a_2;\,a_3;\,a_4)$
provides the regular solution of the hypergeometric differential equation and converges for $\lvert a_4\rvert<1$.  It admits the well-known power series expansion  
\begin{equation}
{}_2F_1(a_1,a_2;\,a_3;\,a_4)
=\sum_{p=0}^{\infty}
\frac{(a_1)_p\,(a_2)_p}{(a_3)_p}\,\frac{a_4^p}{p!}\,,
\end{equation}
where $(a)_p = a\,(a+1)\cdots(a+p-1)$ is the Pochhammer symbol (rising factorial) with $(a)_0=1$.  This function arises ubiquitously in mathematical physics and GR whenever one solves second-order linear differential equations with three regular singular points~\cite{Abra}.

In fact, as $r\to\infty$ the metric function grows without bound, $F(r)\sim r^2/L^2\to+\infty$, reflecting the dominant cosmological (AdS or dS) term rather than approaching a finite constant.  Two further asymptotic regimes merit attention.  First, in the limit $\gamma\to\infty$ with all other parameters fixed, the scalar fluid contribution  
\begin{equation}
z(r)=-\frac{\chi^2\,r^{-n(\beta-1)/\xi}}{\gamma}\;\longrightarrow\;0,
\end{equation}
so that ${}_2F_1(\xi,\xi/(1-\beta);1+\xi/(1-\beta);-z)\to1$ and the entire Murnaghan term vanishes as $\sim\gamma^{-\xi}\,r^2\to0$.  Consequently, the large-$\gamma$ geometry reduces to the usual AdS/dS BH with only mass, gauge and monopole corrections.  Second, when the hypergeometric argument $z\to-\infty$, for example if $\beta>1$ so that $r^{-n(\beta-1)/\xi}$ grows with $r$-one must use the large-$|z|$ expansion  
\begin{equation}
{}_2F_1(a,b;c;z)\sim \frac{\Gamma(c)\,\Gamma(b-a)}{\Gamma(b)\,\Gamma(c-a)}(-z)^{-a}
+\frac{\Gamma(c)\,\Gamma(a-b)}{\Gamma(a)\,\Gamma(c-b)}(-z)^{-b},\,\,a=\xi,\, b=\xi/(1-\beta) ,
\end{equation}
which yields PL decay of the scalar term $\propto r^{-n(\beta-1)}$ (or steeper), ensuring that for $\beta>1$ the Murnaghan fluid contribution becomes subleading at infinity. This yields

\begin{equation}
F_{\rm Mur} (r)\sim  -\frac{2\,(1-\beta)^{\xi}}{n(n-1)\,\chi^{2\xi}}\;
r^{2 + n(\beta-1)}.
\end{equation}
Hence for $\beta\neq1$ the scalar fluid contribution grows (if $\beta>1$) or decays (if $\beta<1$) faster than $r^2$ and does not produce an additional constant $r^2$-term.  The only surviving cosmological-constant-like piece in $F(r)$ remains the bare $\Lambda$-term and if $\epsilon=\tfrac12$—the constant YM shift.  Explicitly,
\begin{equation}
F(r)\sim 1 + \biggl(\frac{1}{L^2}
-\delta_{\epsilon,\tfrac12}\,\frac{(32\pi^2Q^2)^{1/2}}{(n-1)(n-2)\,\Omega_{n-1}}\biggr)\,r^2
+\mathcal O\bigl(r^{2+n(\beta-1)},r^{2-4\epsilon},r^{2-n}\bigr)\,.
\end{equation}
Thus, in the $z\to-\infty$ limit, the Murnaghan SF does not renormalize $\Lambda$: only the bare AdS/dS constant (and the special $\epsilon=1/2$ YM term) survive as true $r^2$-driven cosmological constants.

In Figures \ref{fig3a}--\ref{fig4a}, the behavior of the metric function $ F(r) $ is illustrated across various BH parameter choices for both dS (left panel) and AdS (right panel) configurations. Specifically, Figure \ref{fig3a} shows how $ F(r) $ varies with respect to the Murnaghan fluid parameters $ \gamma $ and $ \beta $. Interestingly, both parameters exhibit qualitatively similar effects on the shape of the metric function. In particular, the variation of either $ \gamma $ or $ \beta $ significantly influences the horizon structure of the BH solution. For instance, when $ \gamma \leq 0.044 $ (or equivalently $ \beta \leq 0.25 $), the function $ F(r) $ reveals the existence of up to three distinct horizons in the dS configuration: an inner horizon, an outer (event) horizon, and a cosmological horizon. In contrast, under the AdS background, the horizon structure reduces to two horizons: the inner and the outer. In contrast, when the parameters exceed the threshold values, i.e., $ \gamma \geq 0.044 $ (or equivalently $ \beta \geq 0.25 $), the metric function $ F(r) $ no longer possesses real positive roots, indicating the absence of horizons and the emergence of a naked singularity. On the other hand, considering the variation of the monopole charge parameter $\eta$ (see Figure \ref{fig4a}), the horizon structure changes significantly. As $\eta$ increases from 0.01 to 0.09, the dS configuration transitions from having three possible horizons to only the inner horizon remaining. In contrast, the AdS configuration exhibits a stable horizon structure that remains essentially unaffected by variations in the GM charge. Similarly, for each spacetime dimension $ n \geq 5 $, with fixed parameter values and a specific choice of the PYM exponent $\epsilon$, the horizon structure in the dS configuration can exhibit up to three horizons. In contrast, under the AdS configuration, the horizon structure consistently features only two horizons: the inner and the outer.

Figure~\ref{emm} illustrates the embedding diagram of the BH spacetime for different values of the dimensionality parameter $n$. The curves demonstrate how the geometry of the equatorial slice evolves as $n$ is varied.  For smaller $n$ the embedding profile bends more sharply near the throat, reflecting stronger curvature effects and a steeper potential barrier.  As $n$ increases the embedding surface becomes progressively smoother and less curved, indicating that higher dimensions dilute the local gravitational distortion and reduce the depth of the embedding well.  This behaviour is consistent with our analytic findings: larger $n$ effectively weakens the short-distance contributions to curvature invariants and softens thermodynamic instabilities.  The embedding diagram therefore provides a clear geometric visualization of how the parameter $n$ governs the local shape of the spacetime and moderates the strength of gravitational backreaction.

\subsection{Curvature singularity analysis}
To probe the fundamental geometric properties of the BH solution, we analyze its curvature invariants, specifically, the Ricci scalar $ \mathcal{R} $ and the Kretschmann scalar $ \mathcal{R}_{\alpha\beta\mu\nu} \mathcal{R}^{\alpha\beta\mu\nu} $. These quantities offer a coordinate-independent means of diagnosing singular behavior and verifying the uniqueness and regularity of the spacetime. Both invariants are derived from the metric function~\eqref{metric} and encapsulate the gravitational response of the system to the matter content and geometry, providing critical insight into the physical viability of the solution. Thus, the two scalars are given, respectively by
\begin{align}
\mathcal{R}
&= \frac{1}{n - 2} \Biggl\{
\frac{2 \left( \frac{1 - \beta}{\gamma} \right)^{\xi}}{
\gamma\, r^{\frac{(\beta - 1)(n - 1)}{\xi}} + \chi^2
}\Bigg( 
\left[
\chi^2\left( \beta(n - 1) + 1 \right) 
+ \gamma n\, r^{\frac{(\beta - 1)(n - 1)}{\xi}}
\right]
\left( 1 + \frac{\chi^2\, r^{- \frac{(\beta - 1)(n - 1)}{\xi}}}{\gamma} \right)^{-\xi}\Bigg) \nonumber\\
&+\, 2n\, \Lambda
+ \frac{8\pi \eta^2\, (n - 2)^2}{r^2}
+ 8^{\epsilon}(n - 4\epsilon) \left( \frac{ \pi^{3 - n} Q^2\, \Gamma\left( \frac{n - 1}{2} \right)^2 }{r^4} \right)^{\epsilon}
\Biggr\},\label{R}\\
\mathcal{K} =&\mathcal{R}_{\alpha\beta\mu\nu}\mathcal{R}^{\alpha\beta\mu\nu}= \frac{2(n-3)(n-2)}{r^4}
\Bigg(
\frac{2r^2 \left(\frac{1 - \beta}{\gamma}\right)^{\xi}}{(n-2)(n-1)}\,
{}_2F_1\!\left(\xi, \frac{\xi}{1-\beta}; \frac{\xi}{1-\beta}+1;
-\,\frac{r^{-\,\frac{(n-1)(\beta - 1)}{\xi}} \chi^2}{\gamma} \right)
 \nonumber\\
&
+\, \frac{16\pi M\,r^{3 - n}}{\omega_{n-2}(n-2)}
+\, \frac{2\Lambda\,r^2}{n^2 - 3n + 2}
+\, \frac{8\pi \eta^2}{n-3}
+\, \frac{r^2\, 32^\epsilon \pi^{2\epsilon} \left(\frac{Q^2}{\omega_{n-2}^2 r^4} \right)^{\!\epsilon}}{(n-2)(n - 4\epsilon - 1)}
\Bigg)^2 \nonumber\\
&+\, \frac{4}{(n-2)^2} \Bigg(
\left(\frac{1 - \beta}{\gamma} \right)^{\xi}
\left(
\frac{
\chi^2 \left(\beta + (\beta - 2)(-n) - 5 \right)
+ \gamma (n - 4) r^{\frac{(n - 1)(\beta - 1)}{\xi}}
}{
\gamma r^{\frac{(n - 1)(\beta - 1)}{\xi}} + \chi^2
}
\right)
\left( 1 + \frac{\chi^2 r^{-\frac{(n - 1)(\beta - 1)}{\xi}}}{\gamma} \right)^{-\xi}
 \nonumber\\
&
-\, \frac{(n - 3)(n - 2)}{n - 1}
{}_2F_1\!\left(\xi, \frac{\xi}{1-\beta}; \frac{\xi}{1-\beta}+1;
-\,\frac{r^{-\,\frac{(n-1)(\beta - 1)}{\xi}} \chi^2}{\gamma} \right)
\Bigg)^2 \nonumber\\
&+\, \frac{8}{n - 2} \Bigg(
\frac{(n - 3)\left(\frac{1 - \beta}{\gamma} \right)^{\xi}}{n - 1}
{}_2F_1\!\left(\xi, \frac{\xi}{1 - \beta}; \frac{\xi}{1 - \beta} + 1;
-\,\frac{r^{-\,\frac{(n - 1)(\beta - 1)}{\xi}} \chi^2}{\gamma} \right)
+ \frac{8\pi M (n - 3) r^{1 - n}}{\omega_{n-2}}
 \nonumber\\
&
-\, \frac{2\Lambda}{n - 1}
+ \frac{32^\epsilon \pi^{2\epsilon} (2\epsilon - 1) \left(\frac{Q^2}{\omega_{n-2}^2 r^4} \right)^{\!\epsilon}}{n - 4\epsilon - 1}
- \left( \frac{1 - \beta}{\gamma} \right)^{\xi} \left( 1 + \frac{\chi^2 r^{-\frac{(n - 1)(\beta - 1)}{\xi}}}{\gamma} \right)^{-\xi}
\Bigg)^2 \nonumber\\[1.5ex]
&-\, \frac{1}{(n - 2)^2} \Bigg(
\frac{32\pi M (n - 3)(n - 2) r^{1 - n}}{\omega_{n-2}}
+ \frac{8\Lambda}{n - 1}
- \frac{4 \cdot 32^\epsilon \pi^{2\epsilon} (8\epsilon^2 - 6\epsilon + 1) \left(\frac{Q^2}{\omega_{n-2}^2 r^4} \right)^{\epsilon}}{-n + 4\epsilon + 1}
\Bigg). \label{RRR}
\end{align}
By carefully analyzing the curvature invariants given in Eqs.~\eqref{R} and~\eqref{RRR}, we confirm that the BH solution defined by the metric function~\eqref{metric} exhibits a curvature singularity. This result holds for any physically admissible choice of parameters in spacetime dimensions $ n > 3 $. The emergence of the singularity is primarily attributed to the divergence of the mass term, the YM charge contribution, the GM configuration, and the Murnaghan distribution sector near the origin. However, in lower-dimensional settings specifically for $ n = 3 $, such divergent behavior can be eliminated, as certain curvature contributions vanish identically. This opens the door to alternative regular geometries, and indeed, various models have been proposed in the literature to construct non-singular BH spacetimes via modified gravity or non-linear electrodynamics (see, e.g.,~\cite{Balart:2014}). In the present analysis, we do not pursue such regularization mechanisms and instead focus on the physically realistic scenario governed by the full metric~\eqref{metric}. A direct evaluation of the Ricci scalar and the Kretschmann scalar near the center $ r = 0 $ confirms the singular nature of the solution, as both invariants diverge in this limit, as the results below show:
\begin{align}
    \lim\limits_{r\to 0} \mathcal{R} &\approx\infty,\qquad
    \lim\limits_{r\to 0}\mathcal{R}_{\alpha\beta\mu\nu}\mathcal{R}^{\alpha\beta\mu\nu} \approx\infty\,.
\end{align}
At large distances, both curvature invariants remain finite, with
\begin{equation}
\lim_{r \to \infty} \mathcal{R} \simeq \frac{2n}{n - 2}\,\Lambda, \qquad
\lim_{r \to \infty} \mathcal{R}_{\alpha\beta\mu\nu} \mathcal{R}^{\alpha\beta\mu\nu} \simeq \frac{8n}{(n - 2)^2(n - 1)}\,\Lambda^2.
\end{equation}
This confirms that the spacetime asymptotically approaches a maximally symmetric (A)dS geometry, with the curvature structure governed entirely by the cosmological constant.

\section{Thermodynamics, Smarr relation and stability}
\label{sec3}
In the framework of gauge/gravity duality, strongly coupled gauge theories are related to weakly coupled string theories. Holographically, information flows bidirectionally: bulk string dynamics encode properties of the boundary gauge theory, while conversely, the gauge theory captures aspects of the bulk. In particular, the AdS/CFT correspondence \cite{Maldacena:1997re,Witten:1998qj} identifies a conformal field theory on the boundary with a higher-dimensional spacetime that asymptotically approaches AdS. Within this duality, the thermodynamic behavior of a BH in AdS space directly reflects the thermal state of its boundary CFT, and the BH’s event horizon encodes the temperature of the dual field theory.

\subsection{Thermodynamics}
In what follows, we investigate the thermodynamic properties of the non-linearly charged AdS BH immersed in a medium governed by the Murnaghan EoS by deriving the Hawking temperature based on the computation of the surface gravity at the event horizon~\cite{Kubiznak:2016qmn}, and verifying both the extended first law of thermodynamics and the generalized Smarr relation. In order to derive the Hawking temperature, we first compute the surface gravity at the event horizon.  The surface gravity is defined by~\cite{Kubiznak:2016qmn}  
\begin{equation}\label{r30}
    \kappa =\sqrt{-\tfrac12\,\nabla_\mu\xi_\nu\,\nabla^\mu\xi^\nu}
    =\tfrac12\,F'\bigl(r_h\bigr)\,,
\end{equation}
where $\xi^\mu = \partial_t$ is the timelike Killing vector of the metric. By recognising the relation $T = \kappa/2\pi$, the Hawking temperature can be expressed directly in terms of the BH parameters, following the resolution of the horizon condition $F(r_h)=0$ for the mass $M$ as outlined below:
\small
    \begin{equation} \label{43}
     T=-\frac{1}{4 \pi  r_h\ (n-2)}\Biggl\{r_h^2 \Bigg(2 \Lambda +8^{\epsilon } \bigg(\frac{\pi ^{3-n} Q^2 \Gamma
   \left(\frac{n-1}{2}\right)^2}{r_h^4}\bigg)^{\epsilon }+2 (1-\beta)^{\xi } \bigg(\chi ^2 r_h^{-\frac{(\beta
   -1) (n-1)}{\xi }}+\gamma\bigg)^{-\xi }\Bigg)-16 \pi  \eta ^2+n \left(8 \pi  \eta ^2-n+5\right)-6\Biggr\},
    \end{equation}
which, under a consistent consideration of the limit $(\gamma \rightarrow \infty,\, \epsilon = 1,\, \beta = 0,\, \eta = 0,\, Q = 0,\, \Lambda = 0)$, yields a Hawking temperature from Eq.~\eqref{43} that reduces to that of the Schwarzschild-Tangherlini spacetime.

Analytically, the dominant small-$r_h$ behavior is governed by the PYM term $8^{\varepsilon}(\pi^{3-n}Q^{2}\Gamma\!\big(\tfrac{n-1}{2}\big)^{2})^{\varepsilon}r_h^{2-4\varepsilon}$; hence, for realistic $\varepsilon\gtrsim 1/4$ one finds 
$T\sim -\tfrac{8^{\varepsilon}A^{\varepsilon}}{4\pi(n-2)}r_h^{1-4\varepsilon}\to -\infty$ as $r_h\to0^{+}$; this generically forbids arbitrarily small horizons and implies at least one positive root of $T$ when the large-$r_h$ asymptotics differ in sign. For large $r_h$ the cosmological term dominates, so $T(r)\sim -\Lambda \,r_h/(2\pi(n-2))$ and for AdS ($\Lambda<0$) $T\to+\infty$ linearly; by continuity, this combination ensures the existence of at least one extremal radius solving $F(r_h)=0$. The parametereter sensitivity is transparent: $Q,\varepsilon$ amplify small all $r_h$ divergence and promote multi-extremal structure, $\eta$ produces a uniform shift of $F(r)$ (moving roots), $(\beta,\gamma,\xi,\chi)$ shape intermediate-radius wiggles, and $n$ changes prefactors and the coefficient $\pi^{3-n}Q^{2}\Gamma\!\big(\tfrac{n-1}{2}\big)^{2}$ (e.g. $\pi^{3-n}Q^{2}\Gamma\!\big(\tfrac{n-1}{2}\big)^{2}=Q^{2}/4$ for $n=4$, $\pi^{3-n}Q^{2}\Gamma\!\big(\tfrac{n-1}{2}\big)^{2}=Q^{2}/\pi^{2}$ for $n=5$).

The thermodynamic entropy of a BH is fundamentally governed by the Bekenstein-Hawking area law~\cite{Bekenstein:1973ur,Gibbons:1977mu}, which posits that the entropy $ S $ is directly proportional to the surface area $ A $ of the event horizon. In a $ n $-dimensional spacetime, this relation is expressed as
\begin{equation}
S = \frac{k_B c^3}{\hbar G} \cdot \frac{A}{4} = \frac{\omega_{n-2}}{4} \, r_h^{n-2}.
\label{eq:BH_entropy}
\end{equation}
This foundational result underscores the holographic principle, suggesting that all thermodynamic information of a BH is encoded on its boundary surface, rather than within its volume.

The corresponding AdS mass $M$ is determined by examining the event horizon at $r = r_h$, where the condition $F(r_h) = 0$ holds. Consequently, the ADM mass can be articulated in relation to the BH parameter space as
\begin{equation}
M=\frac{(2-n)\pi^{\frac{n-3}{2}}r_h^{\,n-3}}{8\Gamma\!\big(\tfrac{n-1}{2}\big)}\Biggl\{\frac{2r_h^{2}D\,{}_2F_{1}(\xi,\frac{\xi}{1-\beta};\frac{\xi}{1-\beta}+1;-d r_h^{k})}{(n-2)(n-1)}+\frac{8\pi\eta^{2}}{n-3}+\frac{r_h^{2}}{n-2}\Big(\frac{2\Lambda}{n-1}+\frac{8^{\varepsilon}A^{\varepsilon}}{\,n-4\varepsilon-1\,}r_h^{-4\varepsilon}\Big)-1\Biggl\} \,, \label{mass}
\end{equation}
with 
\begin{align}
    A=\pi^{3-n}Q^{2}\Gamma\!\big(\tfrac{n-1}{2}\big)^{2},\,\quad
    d=\frac{\chi^2}{\gamma},\,\quad
    D=\left(\frac{1-\beta}{\gamma}\right)^\xi,\,\quad
    k=\frac{(1-\beta)(n-1)}{\xi}.\nonumber
\end{align}
Within our ADM mass \eqref{mass}, several analytic properties are immediate and important for the phase-topology analysis: the closed form exhibits explicit parameter singularities at $n=3$ (the monopole term) and at $n=4\varepsilon+1$ (the PYM coefficient), so these cases require a limiting expansion (they typically produce logarithmic or special-case terms). At small radius $r_h\to0^{+}$, the hypergeometric argument $z=-d r_h^{k}\to0$ (for $\beta<1$) and $ {}_2F_{1}\to1+O(r_h^{k})$, while the algebraic PYM piece scales like $r_h^{2-4\varepsilon}$ inside the bracket and hence contributes to $M$ as $M_{\rm PYM}\sim r_h^{\,n-1-4\varepsilon}$; for parameter ranges with $n-1-4\varepsilon<0$ (typical for $\varepsilon\gtrsim0.25$ and small $n$) this yields a small-$r_h$ divergence and signals either a pathology or the emergence of a minimal admissible horizon radius. At large $r_h$, the hypergeometric and PYM corrections decay as powers, while the cosmological contribution, which is proportional to $r_h^{2}$, dominates inside the bracket, so $M(r_h)\sim r_h^{\,n-1}$, recovering the expected extensive AdS scaling. In particular, for the commonly used benchmarks, one has the simple identifications $A=Q^{2}/4$ for $n=4$ and $A=Q^{2}/\pi^{2}$ for $n=5$, which change numerical prefactors but leave the qualitative small- and large-$r_h$ scalings unchanged. Monotonicity and extrema of $M(r_h)$ follow from $dM/dr_h$ (the derivative combines the outer $r_h^{n-3}$ prefactor and the derivative of the bracket); at large $r_h$ one generically finds $dM/dr_h>0$, while at small $r_h$ the PYM term can render $dM/dr_h$ large or sign-indefinite, thereby connecting to the temperature/heat-capacity analysis via $dM=T\,dS$ and $C=T(dS/dr_h)/(dT/dr_h)$. 


\begin{figure*}[tbh!]
      	\centering{
       \includegraphics[height=7cm,width=8.0cm]{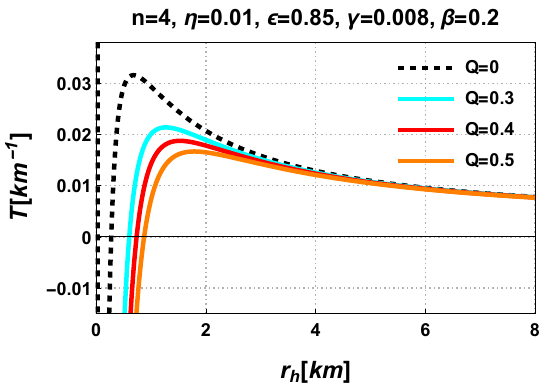} \hspace{1cm}
      	\includegraphics[height=7cm,width=8.0cm]{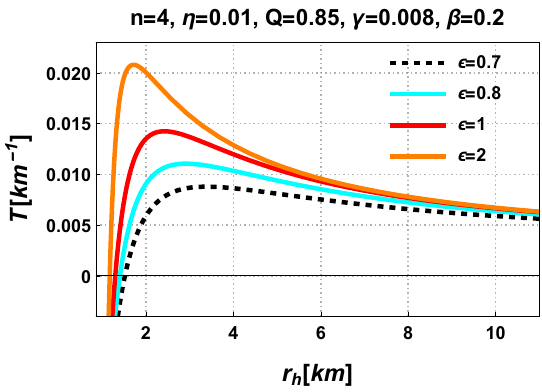} \hspace{1cm}
       \includegraphics[height=7cm,width=8.0cm]{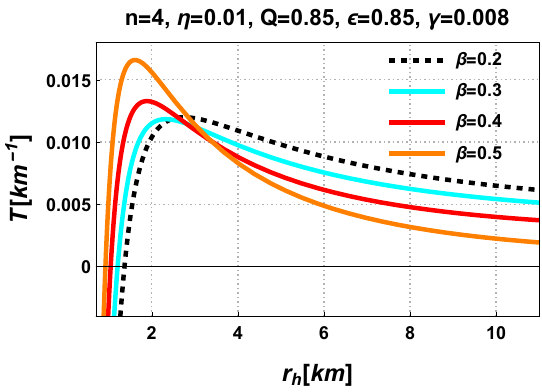}
       \hspace{1cm}
       \includegraphics[height=7cm,width=8.0cm]{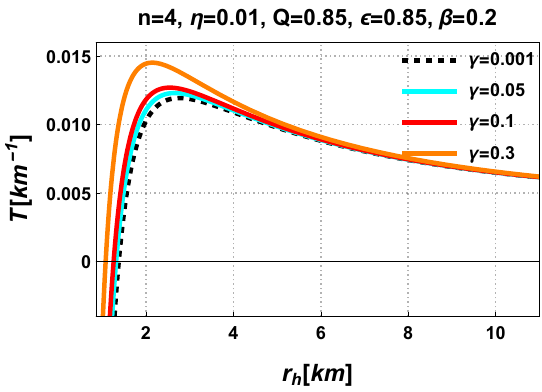}
    \includegraphics[height=7cm,width=8.0cm]{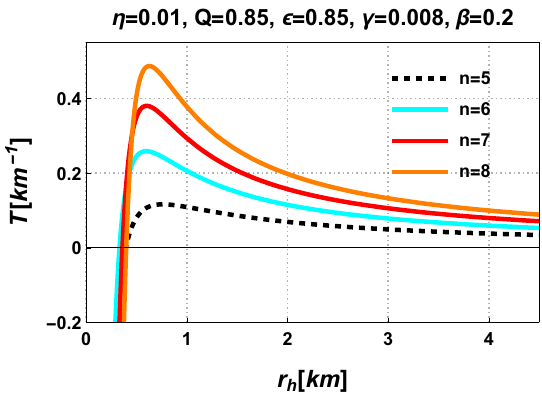}
       \hspace{1cm}
       \includegraphics[height=7cm,width=8.0cm]{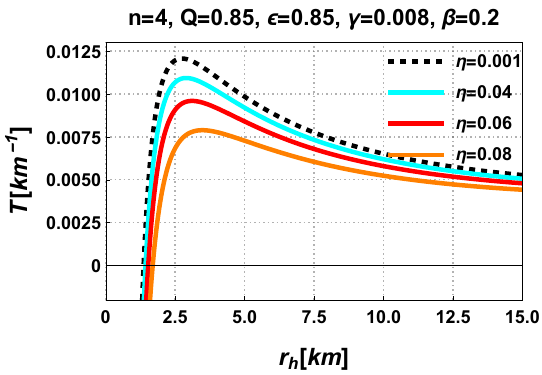}
      }
       
      	\caption{The Hawking temperature is plotted against the event horizon radius $r_h[km]$ for various parameter sets, specifically with $\Lambda$ set to -0.01. } 
    \label{fig5a}
      \end{figure*}

\subsection{ Smarr relation }
From an alternative perspective, one can establish an appropriate thermodynamic framework by differentiating the mass as $M = M(S, P, Q, \eta, \chi)$, which results in the extended first law being expressed in a compact form such that 
\begin{equation}
\mathrm{d}M=T\,\mathrm{d}S+V\,\mathrm{d}P+\Phi\,\mathrm{d}Q+\Psi\,\mathrm{d}\eta+\Theta\,\mathrm{d}\chi,
\end{equation} 
where $T = (\partial_S M)_{P,Q,\eta,\chi}$ denotes the Hawking temperature, $V = (\partial_P M)_{S,Q,\eta,\chi} = \Omega_{n-2}\,r_h^{n-1}/n-1$ represents the thermodynamic (geometric) volume, $\Phi = (\partial_Q M)_{S,P,\eta,\chi}$ signifies the PYM potential, $\Psi = (\partial\eta M)_{S,P,Q,\chi}$ refers to the monopole conjugate, and $\Theta = (\partial\chi M)_{S,P,Q,\eta}$ denotes the Murnaghan-fluid conjugate, as defined in the following manner:
\begin{align}\label{r52}
\Phi&=\left(\frac{\partial M}{\partial Q}\right)_{S,P,\gamma}=\frac{\pi ^{\frac{n-3}{2}} 2^{3 \epsilon -2} \epsilon  r_h^{n-1} \left(\frac{\pi ^{3-n} Q^2 \Gamma
   \left(\frac{n-1}{2}\right)^2}{r_h^4}\right)^{\epsilon }}{Q (-n+4 \epsilon +1) \Gamma \left(\frac{n-1}{2}\right)}, \\
\Psi&=\left(\frac{\partial M}{\partial \eta}\right)_{S,Q,\gamma}=-\frac{2 \eta  (n-2) \pi ^{\frac{n-1}{2}} r_h^{n-3}}{(n-3) \Gamma \left(\frac{n-1}{2}\right)},\\
\Theta&=\left(\frac{\partial M}{\partial \chi}\right)_{S,Q,P}=\frac{\pi ^{\frac{n-3}{2}} \xi  r_h^{n-1} \left(\frac{1-\beta }{\gamma }\right)^{\xi } \left(\left(\frac{\chi ^2 r_h^{-\frac{(\beta -1) (n-1)}{\xi
   }}}{\gamma }+1\right)^{-\xi }-\, _2F_1\left(\xi ,\frac{\xi }{1-\beta };\frac{\xi }{1-\beta }+1;-\frac{r_h^{-\frac{(n-1) (\beta -1)}{\xi }} \chi
   ^2}{\gamma }\right)\right)}{4 (\beta -1) \chi  \Gamma \left(\frac{n+1}{2}\right)}.
\end{align}
Equivalently, one may introduce the horizon-constraint function $F(r_h,M,Q,\eta,P,\chi)=0$ and impose $\mathrm{d}F=0$, solving for $\mathrm{d}M$ to recover the same first law with $T=F'(r_h)/(4\pi)$ and $\mathrm{d}S=(-4\pi\,\partial_M F)^{-1}\mathrm{d}r_h$, and analogous identifications for $\Phi$, $\Psi$, and $\Theta$.  By dimensional scaling $r_h\to\lambda r_h$ one finds $[M]\sim L^{n-3}$, $[S]\sim L^{n-2}$, $[P]\sim L^{-2}$, $[Q]\sim L^{n-2\epsilon}$, $[\eta]\sim L^0$, $[\chi]\sim L^{n(\beta-1)/\xi}$, and Euler’s theorem~\cite{Kastor:2009wy,Altamirano:2014tva}  then yields the generalized Smarr relation  
\begin{equation}
    (n-3)\,M=(n-2)\,T\,S-2\,P\,V+(n-2\epsilon)\,\Phi\,Q
+\frac{n(\beta-1)}{\xi}\,\Theta\,\chi,
\end{equation}
in which the $\Psi\,\eta$ term vanishes by scale invariance, the $-2PV$ term encodes AdS pressure-volume contribution such that $P=-\Lambda/8\pi$, $2\epsilon\,\Phi Q$ captures the PYM non-linearity, and the last term encodes the Murnaghan-fluid scaling. For a specific spacetime dimension such that $n=4$, the Smarr relation reduces to 
\begin{equation}
    M=2TS-2PV+2\epsilon\,\Phi Q+\tfrac{3(\beta-1)}{\xi}\Theta\chi .
\end{equation}
While for $n=5$, the Smarr formula takes the following form:\begin{equation}
    2M=3TS-2PV+2\epsilon\,\Phi Q+\tfrac{4(\beta-1)}{\xi}\Theta\chi.
\end{equation}
 In the classical limit, where $ Q = \ 0 $ and $ \chi = 0 $, the thermodynamic quantities derived herein coincide with those of the AdS Schwarzschild-Tangherlini BH~\cite{Kumar:2020}. This concordance underscores the robustness of the extended framework, which adeptly encompasses both charged and neutral configurations, thereby providing a comprehensive foundation for investigating the phase structure and critical phenomena of BHs immersed in a monopole-fluid Murnaghan EoS environment.

\begin{figure*}[tbh!]
      	\centering{
       \includegraphics[height=6cm,width=5.2cm]{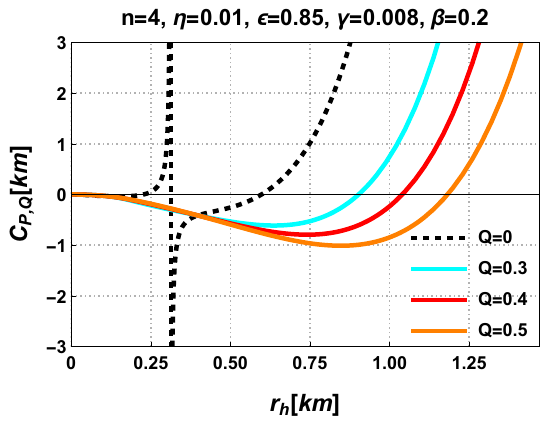} \hspace{3mm}
      \includegraphics[height=6cm,width=5.5cm]{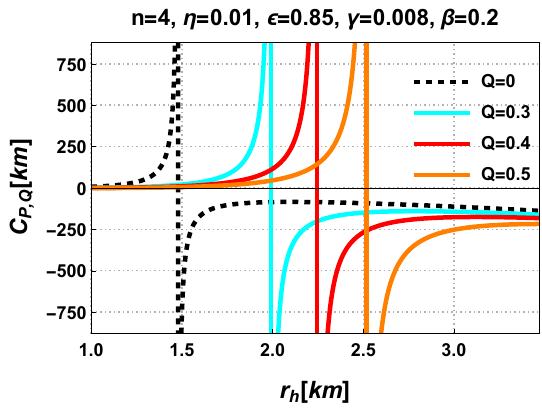} \hspace{3mm}
       \includegraphics[height=6cm,width=5.95cm]{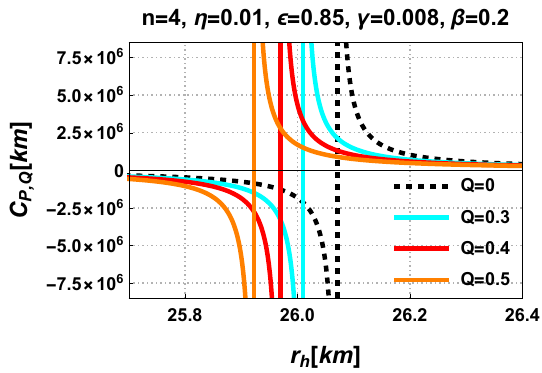}}
       \caption{The heat capacity variation versus event horizon radius $r_h$ for various values of the PYM charge $Q$, specifically with $\Lambda$ set to -0.01.} 
    \label{fig6a}
      \end{figure*}

\begin{figure*}[tbh!]
      	\centering{\includegraphics[height=6cm,width=5.2cm]{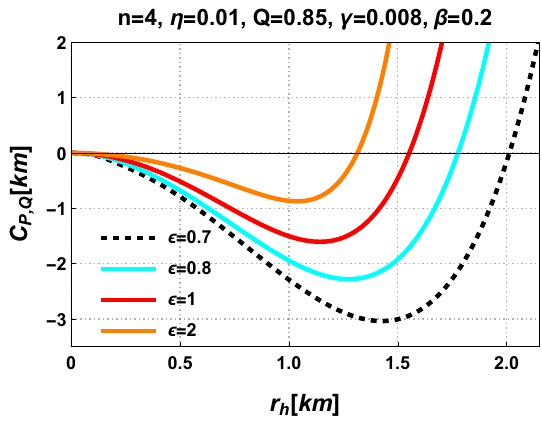} \hspace{3mm}
      \includegraphics[height=6cm,width=5.5cm]{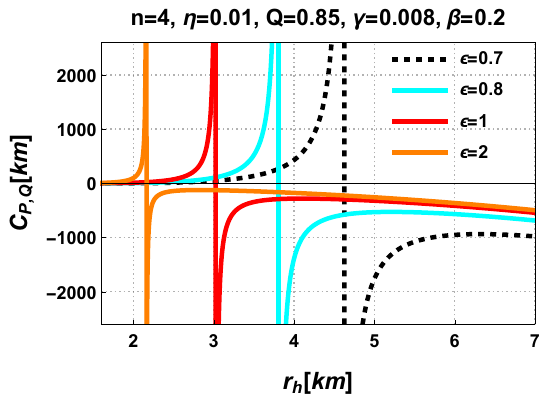} \hspace{3mm}
       \includegraphics[height=6cm,width=5.95cm]{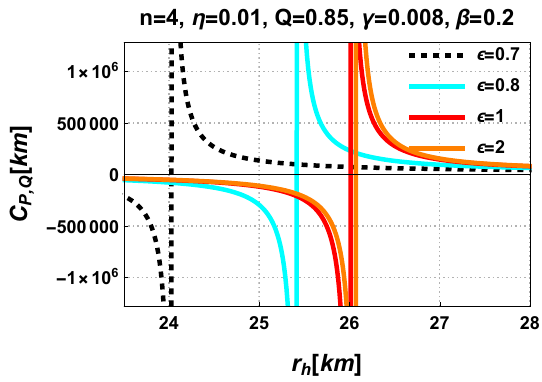}
     }
       \caption{The heat capacity variation versus event horizon radius $r_h$ for various values of the PYM nonlinearity $\epsilon$, specifically with $\Lambda$ set to -0.01.} 
    \label{fig6b}
      \end{figure*}
       
       \begin{figure*}[tbh!]
      	\centering{\includegraphics[height=6cm,width=5.2cm]{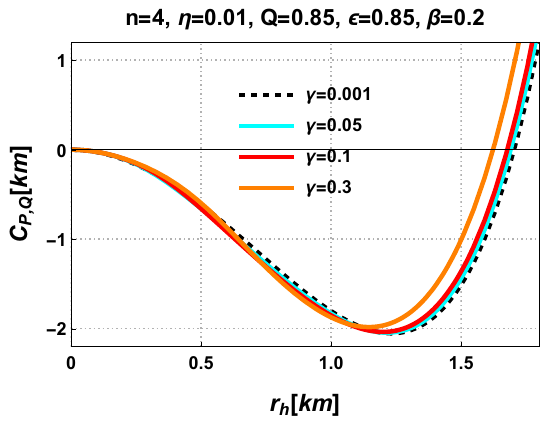} \hspace{3mm}
      \includegraphics[height=6cm,width=5.5cm]{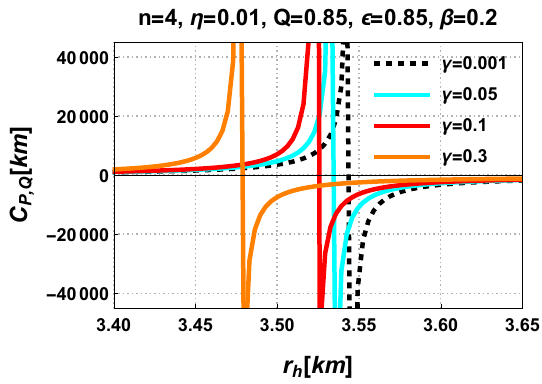} \hspace{3mm}
       \includegraphics[height=6cm,width=5.95cm]{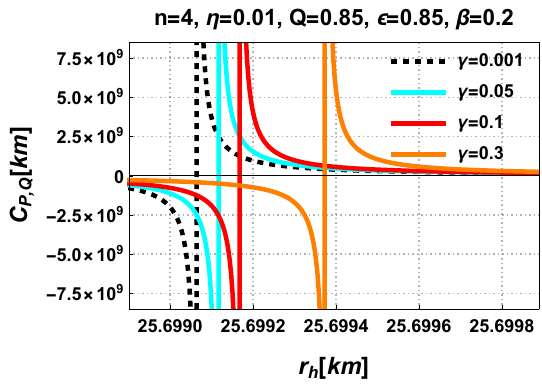}
      }
       \caption{The heat capacity variation versus event horizon radius $r_h$ for various values of the Murnaghan modulus parameter $\gamma$, specifically with $\Lambda$ set to -0.01.} 
    \label{fig6c}
      \end{figure*}
       
         \begin{figure*}[tbh!]
      	\centering{\includegraphics[height=6cm,width=5.2cm]{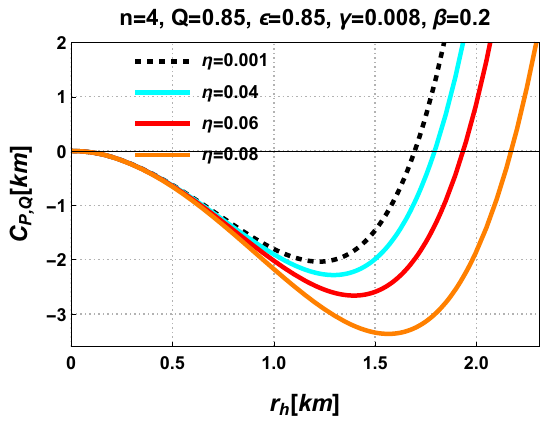} \hspace{3mm}
      \includegraphics[height=6cm,width=5.5cm]{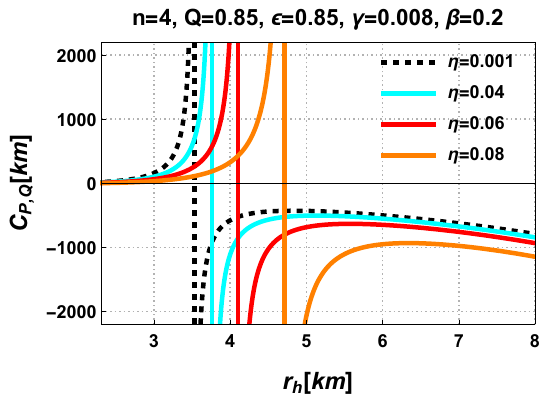} \hspace{3mm}
       \includegraphics[height=6cm,width=5.95cm]{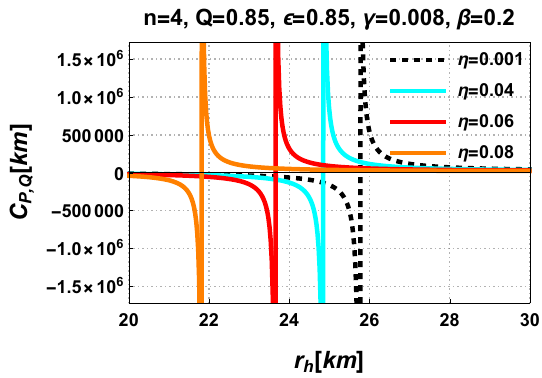}
       }
       \caption{The heat capacity variation versus event horizon radius $r_h$ for various values of the global-monopole charge $\eta$, specifically with $\Lambda$ set to -0.01.} 
    \label{fig6d}
      \end{figure*}

         \begin{figure*}[tbh!]
      	\centering{\includegraphics[height=6cm,width=5.2cm]{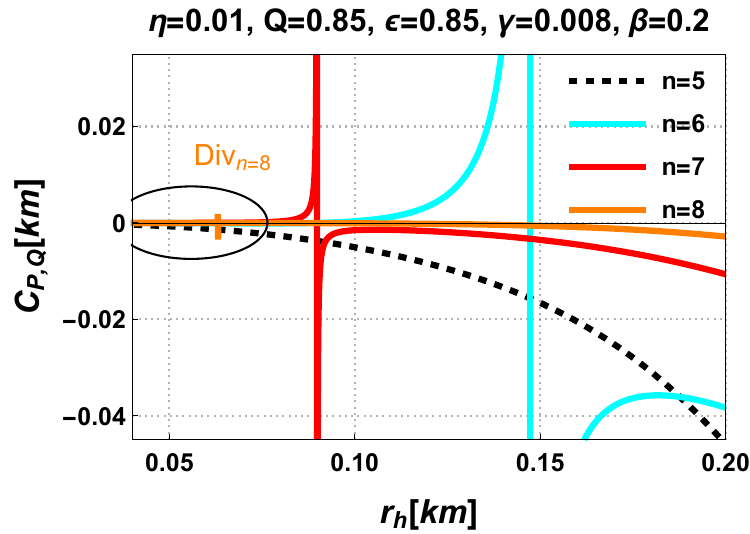} \hspace{3mm}
      \includegraphics[height=6cm,width=5.5cm]{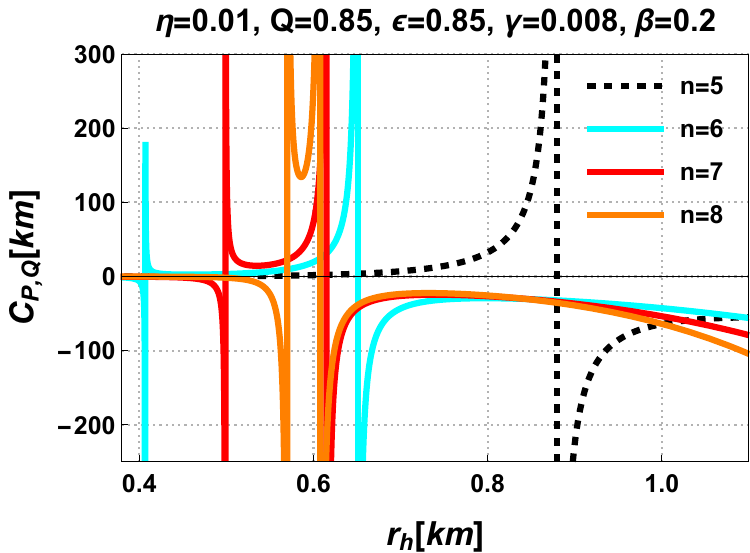} \hspace{3mm}
       \includegraphics[height=6cm,width=5.95cm]{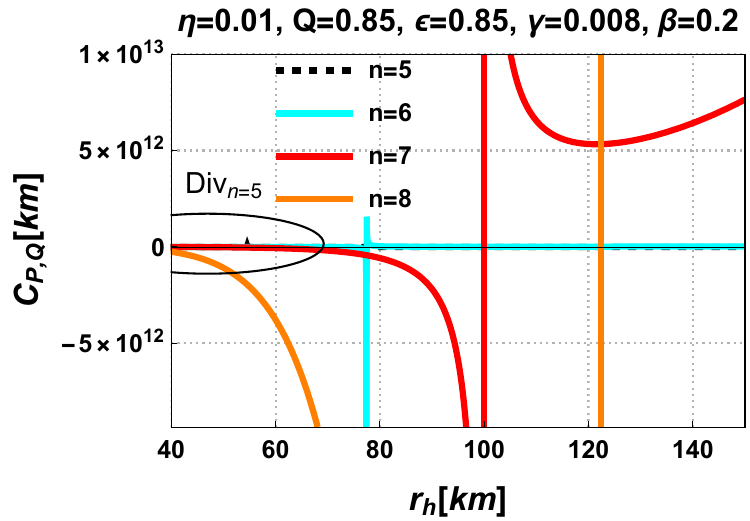}
       
      }
       
      	\caption{The heat capacity variation versus event horizon radius $r_h$ for various spacetime dimensions $n=5,6,7$ and $8$, specifically with $\Lambda$ set to -0.01.} 
    \label{fig6e}
      \end{figure*}

      \begin{figure*}[tbh!]
      	\centering{
       
       \includegraphics[height=7cm,width=8.0cm]{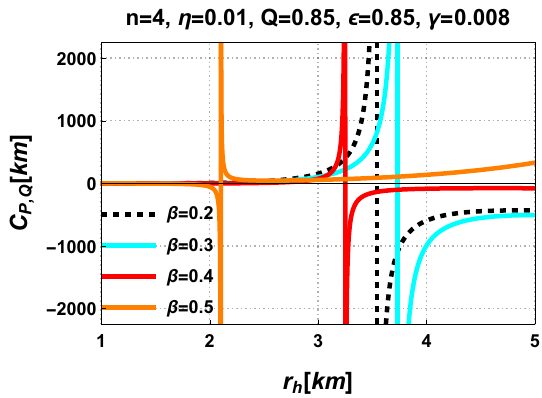} \hspace{3mm}
      \includegraphics[height=7cm,width=8.0cm]{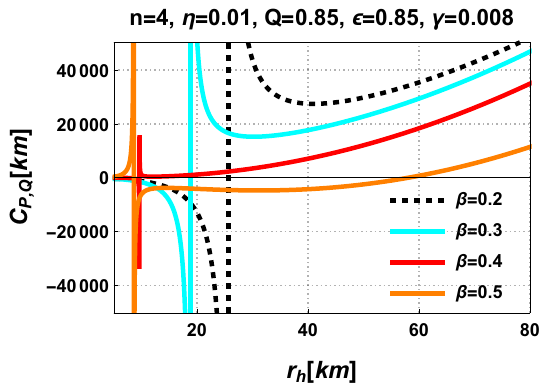} 
      }
       
      	\caption{The heat capacity variation versus event horizon radius $r_h$ for various values of the linear (barotropic) coefficient $\beta$, , specifically with $\Lambda$ set to -0.01..} 
    \label{fig6f}
      \end{figure*}

\subsection{Local stability analysis}
Within the canonical ensemble framework, heat capacity is a crucial thermodynamic quantity that characterises the thermal state of BHs. It provides three key and compelling insights. First, the presence of discontinuities in the heat capacity signals potential thermal phase transitions within the system. Second, the sign of the heat capacity serves as an indicator of thermal stability: a positive value corresponds to a stable thermal equilibrium, whereas a negative value denotes instability~\cite{Sahabandu:2005ma, Cai:2003kt}. Third, the roots of the heat capacity mark critical points where the sign changes, delineating transitions between stable and unstable phases or defining boundary states.

The following analysis focuses on precisely calculating the heat capacity at constant $\lambda$  (pressure $P$) and charge $Q$. This quantity captures the system’s response to thermal fluctuations and is derived from the relevant thermodynamic 
potentials such that~\cite{Kubiznak:2016qmn}
\begin{equation}
C_{P,Q} =T\left(\frac{\partial T}{\partial S}\right)_{P,Q}^{-1}\,.\label{cca}
\end{equation}
Alternatively, in terms of the BH parameter space, the corresponding heat capacity is expressed as:
\begin{align}\label{ch}
C_{P,Q} &= 
\frac{1}{2\,\Gamma\!\bigl(\tfrac{n-1}{2}\bigr)}\,
\frac{\mathcal{N}}{\mathcal{D}}\,,
\end{align}
with
\begin{align}
\mathcal{N} &=
(n-2)\,\pi^{\frac{n-1}{2}}\,
r_h^{\,n-2}\,
\bigl(\gamma\,r_h^{\frac{(\beta-1)(n-1)}{\xi}} + \chi^2\bigr)\,
\Biggl[
  (n-2)\bigl(-8\pi\eta^2 + n-3\bigr)\,
  \Bigl(\tfrac{\chi^2\,r_h^{-\frac{(\beta-1)(n-1)}{\xi}}}{\gamma}+1\Bigr)^{\!\xi}\nonumber\\
&\qquad\quad
  -\,r^2\Bigl(
      2\Bigl(\tfrac{1-\beta}{\gamma}\Bigr)^{\!\xi}
    +\bigl(2\Lambda + 8^{\epsilon}\bigl(\tfrac{\pi^{3-n}Q^2\Gamma(\tfrac{n-1}{2})^2}{r_h^4}\bigr)^{\!\epsilon}\bigr)\,
      \Bigl(\tfrac{\chi^2\,r_h^{-\frac{(\beta-1)(n-1)}{\xi}}}{\gamma}+1\Bigr)^{\!\xi}
  \Bigr)
\Biggr],\nonumber\\[6pt]
\mathcal{D} &=
\gamma\,r_h^{\frac{(\beta-1)(n-1)}{\xi}+2}\,
\Biggl[
  \bigl(8^{\epsilon}(4\epsilon-1)\bigl(\tfrac{\pi^{3-n}Q^2\Gamma(\tfrac{n-1}{2})^2}{r_h^4}\bigr)^{\!\epsilon}
  -2\Lambda\bigr)\,
  \Bigl(\tfrac{\chi^2\,r_h^{-\frac{(\beta-1)(n-1)}{\xi}}}{\gamma}+1\Bigr)^{\!\xi}
  -2\Bigl(\tfrac{1-\beta}{\gamma}\Bigr)^{\!\xi}
\Biggr]\nonumber\\
&\quad
+\,r_h^2\chi^2\,
\Biggl[
  -4\Bigl(\tfrac{1-\beta}{\gamma}\Bigr)^{\!\xi}
  -2\bigl((\beta-1)n-\beta\bigr)\Bigl(\tfrac{1-\beta}{\gamma}\Bigr)^{\!\xi}
  +\,\bigl(8^{\epsilon}(4\epsilon-1)\bigl(\tfrac{\pi^{3-n}Q^2\Gamma(\tfrac{n-1}{2})^2}{r_h^4}\bigr)^{\!\epsilon}
  -2\Lambda\bigr)\,
  \Bigl(\tfrac{\chi^2\,r_h^{-\frac{(\beta-1)(n-1)}{\xi}}}{\gamma}+1\Bigr)^{\!\xi}
\Biggr]\nonumber\\
&\quad
+\,\gamma\,(2-n)\,\bigl(-8\pi\eta^2 + n-3\bigr)\,
   r_h^{\frac{(\beta-1)(n-1)}{\xi}}\,
   \Bigl(\tfrac{\chi^2\,r_h^{-\frac{(\beta-1)(n-1)}{\xi}}}{\gamma}+1\Bigr)^{\!\xi}
- (n-2)\,\chi^2\,\bigl(-8\pi\eta^2 + n-3\bigr)\,
   \Bigl(\tfrac{\chi^2\,r_h^{-\frac{(\beta-1)(n-1)}{\xi}}}{\gamma}+1\Bigr)^{\!\xi}\,.\nonumber
\end{align}
Analytically speaking this rational structure makes the diagnostics transparent: $\mathcal{N}=0$ yields $C=0$ (sign change of stability) while $\mathcal{D}=0$ gives $|C|\to\infty$ (divergent heat capacity, typically a second-order critical point). At small horizon radius the dominant behavior is controlled by the PYM algebraic power $r_h^{-4\varepsilon}$ and by the Murnaghan amplification $(1+\chi^{2}r^{-k}/\gamma)^{\xi}\sim(\chi^{2}/\gamma)^{\xi}r^{-p}$ where $k=(1-\beta)(n-1)/\xi$ and $p=k\xi=(1-\beta)(n-1)$, so that the leading small--$r_h$ scalings of numerator and denominator pick up large negative powers; consequently for moderate-to-large $\varepsilon$ and/or large $\xi$ the denominator can vanish or diverge faster than the numerator, producing either vanishing $C$ or divergent points near $r_h\to0$ and thus an excluded small-radius region. In the large--$r_h$ regime the non-linear Murnaghan factor tends to unity and the PYM corrections decay, so the cosmological/Murnaghan balance dominates and one finds the asymptotic PL $C_{P,Q}\sim r_h^{\,n-2-k}$, i.e. the Murnaghan combination $k$ controls the large-radius growth or suppression of $C$. Physically, the PYM sector $(Q,\varepsilon)$ determines the intensity of singular behavior at small values of $r_h$. The parameters of Murnaghan EoS $(\chi, \xi, \beta)$ govern both the amplification at small values of $r_h$ through $p$ and the exponent at large values of $r_h$ via $k$. This regulation is crucial as it shapes the intermediate phases and stability states. The GM $\eta$ manifests as a uniform shift in the interval combinations, thereby translating roots and divergent points without altering the asymptotic exponents.

\subsection{Thermodynamic analysis of the Hawking temperature and heat capacity}

The Hawking temperature $T(r_h)$ displayed in Figure \ref{fig5a} — as given explicitly by Eq.~~\eqref{43} encapsulating the competing influences of the PYM charge, the GM parameter, and the Murnaghan scalar gas. Analytically, $T$ is determined by the surface gravity through $T=F'(r_h)/(4\pi)$ and therefore inherits both additive and multiplicative modifications from the monopole contribution and the non-linear PYM sector; the scalar gas enters through backreaction terms controlled by the parameters $(\gamma,\beta,\chi,\xi)$. As a consequence, $T(r_h)$ generically departs from a simple monotonic Schwarzschild–(A)dS profile: for physically relevant parameter regimes the temperature develops local extrema at small-to-moderate horizon radii, whose positions and depths shift systematically when one varies $Q$, $\epsilon$, $\gamma$, $\beta$ or $\eta$. Physically, these extrema mark the boundaries between distinct thermodynamic branches (`small'', `intermediate'' and `large'' BHs) and are the direct origin of the swallowtail coexistence structure observed in the off-shell free energy; when an extremum approaches $T=0$ it additionally signals an endpoint of a branch and the potential emergence of extremal or naked configurations which must be excluded by the horizon-existence conditions.

The heat capacity at constant pressure and charge, $C_{P,Q}(r_h)$, displayed in Figures \ref{fig6a}--\ref{fig6f}, admits the rational structure displayed in Eq.~~\eqref{cca}, namely
$C_{P,Q}(r_h)=\mathcal{N}(r_h)/\mathcal{D}(r_h),$
with $
\mathcal{N}$ and $\mathcal{D}$ denoting the numerator and denominator functions. This decomposition confers immediate physical interpretation: zeros of $\mathcal{N}$ correspond to vanishing heat capacity and demarcate thermodynamic limitation points (where a branch terminates or the sign of $C$ changes), whereas zeros of $\mathcal{D}$ produce poles in $C$ that indicate genuine second-order (continuous) phase transitions. The numerical plots therefore show alternating bands of positive and negative $C$ — the former indicating local thermodynamic stability and the latter instability — interspersed with divergence loci that are exceedingly sensitive to the microscopic parameters. In particular, increasing the PYM charge $Q$ typically depresses $T$ and shifts both zeros and divergent points to larger $r_h$, thereby enlarging the radius-space where small BHs become thermodynamically stable. Conversely, increasing the scalar parameters (reducing $\gamma$ or tuning $\beta$ appropriately) introduces additional small-$r_h$ structure, often creating new sign changes in $C$ and additional critical loci.

The roles of the PYM non-linearity exponent $\epsilon$ and the GM parameter $\eta$ deserve focused emphasis. Lowering $\epsilon$ amplifies deviations from Maxwell-like scaling and sharpens small-radius features, making divergent points in $C$ more significant and multiplying the number of stability aspect; thus $\epsilon$ acts as a control parameter for the microscopic stiffness of the PYM medium. The monopole parameter $\eta$ predominantly produces near-uniform vertical shifts of $T(r_h)$ and, through this mechanism, of $C(r_h)$: as $\eta$ grows the multivaluedness of the thermodynamic functions is negligible and the coexistence swallowtail can eventually disappear, i.e. sufficiently large monopole charge stabilizes the system and converts erstwhile first-order transitions into stable system (without any phase transition). Dimensionally, higher spacetime dimension $n$ modifies the PL scaling of the entropy and gauge contributions; empirically one observes that increasing $n$ tends to decreases multi-extremal behavior and can replace sharp first-order behavior by a stable thermodynamic without any phase transition point, reflecting altered relative scaling between gravitational and gauge/scalar energy densities.

In conclusion, the combined PYM, GM and the Murnaghan-gas system realises a versatile thermodynamic phase structure: non-monotonic temperature profiles engender multiple thermodynamic branches, divergent points of the heat capacity highlight a second-order critical phase transition and coincide with the swallowtail coexisting phases in the free energy, while monopole and scalar parameters act as effective knobs that tune or shrinking multivaluedness.

The plotted curves display the Hawking temperature $T$ as a function of the event horizon radius $r_{h}$ for representative choices of the model parameters (cosmological constant $\Lambda=-0.01$ in the plotted slices of Figure \ref{fig5a}). The principal features are the non-monotonic behavior of $T(r_{h})$ and the existence of one or more local extrema at small-to-intermediate radii. Physically, each local maximum/minimum marks the boundary between thermodynamic branches (small/intermediate/large BHs) and anticipates the presence of phase-structure in the free energy. From the plotted trends we infer that increasing the YM charge $Q$ and enhancing the PYM nonlinearity $\epsilon$ systematically lower the temperature at small $r_{h}$ and push extrema to larger radii, while the monopole parameter $\eta$ produces an overall horizontal/vertical translation of the extrema without generating new small-radius structure. These behaviors justify and motivate the subsequent heat-capacity analysis: the radii at which $T(r_{h})$ exhibits maxima or minima coincide with changes in sign or singularities of $C_{P,Q}$, and therefore with local thermodynamic stability boundaries.


Figure \ref{fig6a} shows the heat capacity $C_{P,Q}$ plotted versus $r_{h}$ while varying the PYM charge $Q$. The most striking effect is the systematic displacement of singular point (vertical divergences of $C_{P,Q}$) toward larger horizon radii as $Q$ increases. In addition, the size and sharpness of the surrounding peaks grows with $Q$, indicating progressively stronger second-order critical behavior. Moreover, the intermediate region defined by the two divergent points of the heat capacity, which exhibit the intermediate BH phase, decreases in size whenever the PYM charge increases. This remarkable observation shows that the unstable BH state $(C<
0)$ is dominant in the intermediate phase when the PYM charge is low.  Consequently, BH quickly becomes stable $(C>0)$ when $Q$ is significant. Furthermore, for the neutral case where $Q=0$, the BH thermodynamics system engenders an interesting succession of critical behaviors, namely a triple phase transition (reentrant phase transition), without undergoing any kind of continuous sign change generating the physical limitation point $C=0$. This figure therefore quantifies the role of gauge charge as a control parameter that both shifts and amplifies critical loci; the observed trends support the identification of $Q$ as a tunable knob for the existence and scale of phase transitions.


The family of curves in Figure \ref{fig6b} examines the influence of the PYM nonlinearity index $\epsilon$ on $C_{P,Q}(r_{h})$. Increasing $\epsilon$ sharpens and heightens the small-radius divergences, producing singularities on the heat capacity (divergent points) that are absent or much milder when $\epsilon$ is small. This sharpening reflects the analytic small-$r_{h}$ scaling induced by non-linear gauge dynamics: nonlinearity increases the effective curvature contribution at short scales and therefore magnifies thermodynamic singularities. More concretely, the modification of the PYM nonlinearity index $\epsilon$ reveals fascinating characteristics in the behavior of heat capacity. Firstly, adjustments to $\epsilon$ at small values of $r_h$ result in a shift in the loci of the physical limitation points $(C=0\Leftrightarrow T=0)$. In particular, the modification affects both small-intermediate and large BH sizes. Particularly, when the PYM nonlinearity index $\epsilon$ is low, the size of the intermediate region that displays an intermediate BH phase is quite limited. Consequently, the transition to a larger BH size with a stable state occurs rapidly. 
Practically, Figure \ref{fig6b} demonstrates that non-linear YM dynamics can create or amplify critical behavior localized at small horizons; consequently, parameter ranges with large $\epsilon$ must be treated with care and, when physically relevant, may require additional regularization or quantum corrections.


Figure \ref{fig6c} presents $C_{P,Q}(r_{h})$ for different values of the Murnaghan modulus $\gamma$. More concretely, the modification of the Murnaghan modulus $\gamma$ reveals fascinating characteristics in the behavior of heat capacity. Firstly, adjustments to $\gamma$ at small values of $r_h$ result in a shift in the loci of the physical limitation points $(C=0\Leftrightarrow T=0)$. In particular, the modification affects both small-intermediate and large BH sizes. In particular, when the Murnaghan modulus $\gamma$ is small, the size of the intermediate region shows that the intermediate BH phase is quite limited. Consequently, the transition to a larger BH size with a stable state occurs rapidly. In other words,  the dominant effect of decreasing $\gamma$ (stronger scalar-fluid backreaction for fixed other parameters) is the appearance of additional divergent points and zeros at intermediate $r_{h}$ and an increase in the complexity of the intermediate radius structure. Conversely, larger $\gamma$ tends to smooth the heat capacity profile and to shift or remove intermediate divergent points. This behavior confirms that the scalar gas backreaction is not merely perturbative: the Murnaghan non-linear pressure term actively sculpts the thermodynamic landscape and can induce new transition loci at intermediate scales that do not originate from the gauge sector or GM.


In Figure \ref{fig6d} we vary the global-monopole parameter $\eta$ and observe a largely coherent translation of the $C_{P,Q}$ curves: divergent points and zeros are displaced collectively and the amplitude of large-radius divergences increases with $\eta$. It is observed that the intermediate region defined by the two divergent points of the heat capacity, which exhibit the intermediate BH phase, decreases in size whenever the GM charge increases. This remarkable observation shows that the unstable BH state $(C<
0)$ is dominant in the intermediate phase when the GM charge is low.  Consequently, BH quickly becomes stable $(C>0)$ when $\eta$ is significant.
The monopole therefore behaves as a geometric tuning parameter (via the solid-angle deficit it induces) which shifts the location of thermodynamic features without generating novel small-scale singularities. From a physical viewpoint this implies that the monopole changes the effective horizon-radius mapping and the observational scale of phase transitions, but it does not, by itself, generate additional local instabilities at small $r_{h}$.


Figure \ref{fig6e} compares $C_{P,Q}(r_{h})$ across several spacetime dimensions $n$. The principal trend is a dimensional softening of the most extreme small-radius behavior: increasing $n$ reduces the height of small-$r_{h}$ peaks and shifts critical loci, while altering the large-$r_{h}$ asymptotic scaling in accordance with the entropy law $S\propto r_{h}^{\,n-2}$. In short, geometric dilution in higher dimensions competes with matter backreaction and can mitigate thermodynamic singularities; this observation confirms that the qualitative multi-branch phase structure is robust across dimensions, though its quantitative manifestation is strongly $n$-dependent. A concrete observation of the heat capacity behavior across different spacetime dimensions reveals additional features beyond the standard four-dimensional spacetime $(n=4)$. Specifically, multi-branches have appeared, indicating the so-called reentrant phase transition. Regarding counting, the five-dimensional case shows two discontinuous sign changes, one of which appears at a quantum scale, leading to a first-order phase transition between small and large BHs. In dimensions from six to eight, the BH system experiences multi-branching with abrupt sign changes, indicating more than four phase transitions, including reentrant phase transitions. 


Figure \ref{fig6f} displays $C_{P,Q}(r_{h})$ for varying barotropic coefficient $\beta$ of the generalized polytropic (Murnaghan) EoS. Increasing $\beta$ shifts Murnaghan-dominated features to smaller radii and can reduce the number of visible divergent points in the plotted graph, while small $\beta$ allows additional intermediate extrema and divergent points to emerge. This indicates a significant degeneracy in phenomenology: different combinations of $\beta$ and the other Murnaghan parameters can reveal or conceal thermodynamic transitions within the same observational radius range. Consequently, constraining the scalar gas EoS is essential to lift degeneracies, when compared to putative observational probes. In more concrete terms, the variations of parameter $\beta$, in contrast to the other Murnaghan parameters, remove the physical limitation point that evolves through a continuous change in sign.

\section{Thermodynamic topology}
\label{sec4}
Recent advancements in the study of BH thermodynamics have highlighted the importance of thermodynamic topology as a tool to investigate the complex phase structure of BHs. Initially, topological methods were used to investigate phenomena such as light rings and time-like circular orbits within BH spacetimes \cite{A,A2}. The application of topological techniques in BH thermodynamics was first demonstrated in \cite{28}, drawing on concepts from Duan's ($\phi$-map) method \cite{d1,d2} related to relativistic particle systems. The application of topological techniques in BH thermodynamics was first introduced in \cite{28}, drawing inspiration from Duan's earlier work \cite{d1,d2} in relativistic particle systems. This approach focuses on topological defects, which are represented by the zero points of a vector field and correspond to the critical points of the system. These zero points serve as indicators for phase transitions and can be classified on the basis of their number of windings. This classification allows BH systems to be assigned to distinct topological classes on the basis of their shared thermodynamic properties.

In this study, we adopt the topological technique employed in \cite{29}, which is specifically dedicated to investigating BH thermodynamics. By employing the off-shell free-energy method, we treat BHs as topological defects within their thermodynamic framework. This method offers a comprehensive insight into the local and global topological aspects of BHs, drawing on winding numbers to outline their topological charge and stability. The stability of a BH can be inferred from the sign of its winding number, highlighting the important relationship between thermodynamic topology and BH stability. This method has garnered considerable interest and has been applied to a wide range of BH systems in various gravitational settings \cite{A,A2}.

To investigate the thermodynamic topology and thereby unveil the corresponding phase structure within the framework of the Murnaghan EoS, incorporating non-linear YM and GM charges coupled to the gravity sector, we extend the parameter space to include the Barrow deformation parameter $\Delta$. In this context, we adopt Barrow's proposal, in which the BH horizon exhibits a fractal geometry that enhances its surface area. According to Barrow's modification, the entropy is expressed as
\cite{barrow2,barrow3,leon1,Luciano1}
\begin{equation}
    S=\left(\frac{A}{A_0}\right)^{1+\frac{\Delta}{2}}.
\end{equation}
The next stage consists of constructing the thermodynamic structure based on the Barrow entropy. We begin by establishing the relation between the radius of the horizon and the Barrow entropy through the Bekenstein--Hawking entropy $S_{\mathrm{BH}}$, expressed as
\begin{equation}
   S = \left(\frac{\omega_{n-2}}{4}\,r_h^{n-2}\right)^{1+\frac{\Delta}{2}}\,, 
\end{equation}
where the reciprocal relation can be expressed in terms of the horizon radius as
\begin{equation}\label{rbar}
    r_h=\left(\frac{4}{\omega_{n-2}}S^{\frac{2}{1+\Delta}}\right)^{\frac{1}{D-2}}.
\end{equation}

The overall form of the generalized off-shell free energy was introduced in \cite{29,york} and is defined by
\begin{equation}
\mathcal{F} = E - \frac{S}{\tau}.\label{offshell}
\end{equation}
Here $E$ denotes the BH energy (equivalently its mass $M$, and $S$ its entropy. The parameter $\tau$ is a relaxation timescale, and may be interpreted as the inverse of the local equilibrium temperature defined on a bounding shell surrounding the BH. From this generalized off-shell free energy, one then defines the Duan--$\phi$ vector field as in Ref.~\cite{29}:
\begin{equation}
\phi = \left(\phi^r, \phi^\Theta \right) = \left(\frac{\partial \mathcal{F}}{\partial S}, -\cot \Theta \, \csc \Theta \right).
\label{phi}
\end{equation}
The zeros of the Duan--$\phi$ vector field correspond to the critical points of the BH solutions. In particular, these zeros occur at
\begin{equation}
(\tau,\Theta)=\Bigl(\tfrac{1}{T},\,\tfrac{\pi}{2}\Bigr),
\end{equation}
where $T$ is the BH equilibrium temperature defined on the bounding cavity (or shell). Each such zero (defect) carries a topological charge which is computed using the Duan $\phi$-mapping method. Introducing the unit vector
\begin{equation}
n^a=\frac{\phi^a}{\lVert\boldsymbol{\phi}\rVert},\qquad 
\lVert\boldsymbol{\phi}\rVert=\sqrt{(\phi^1)^2+(\phi^2)^2},
\end{equation}
(with $a=1,2$ and the parameter-space coordinates $(S,\Theta)$), one has the normalization and orthogonality relations
\begin{equation}
n^a n^a = 1,
\qquad
n^a\partial_\nu n^a = 0\qquad(\nu=S,\Theta),
\end{equation}
which expresses that $n^a$ is unit normalized and its derivative is orthogonal to itself.
\begin{figure*}[tbh!]
      	\centering{
       \includegraphics[height=6.8cm,width=7.5cm]{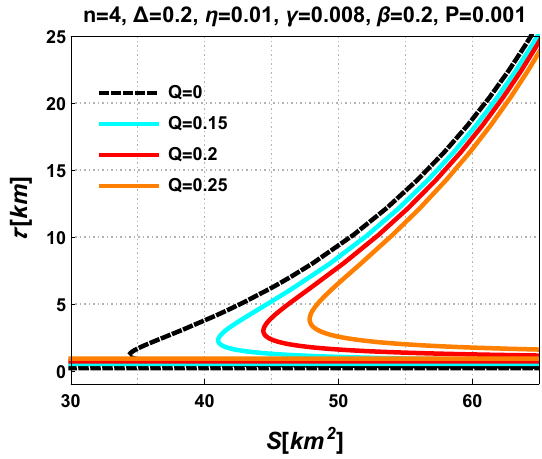} \hspace{1cm}
      	\includegraphics[height=6.8cm,width=7.5cm]{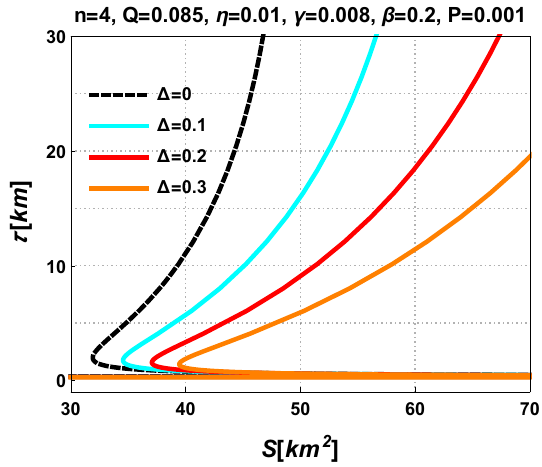} \hspace{1cm}
       \includegraphics[height=6.8cm,width=7.5cm]{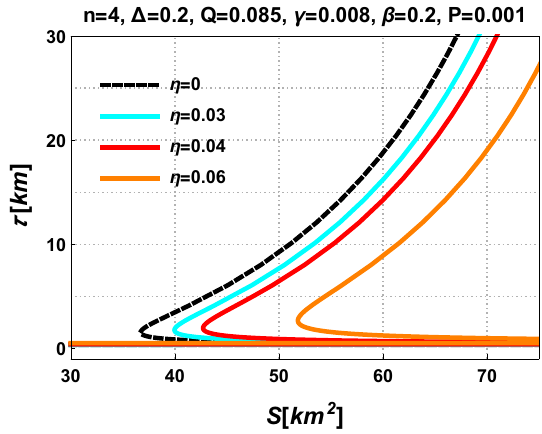}
       \hspace{1cm}
       \includegraphics[height=6.8cm,width=7.5cm]{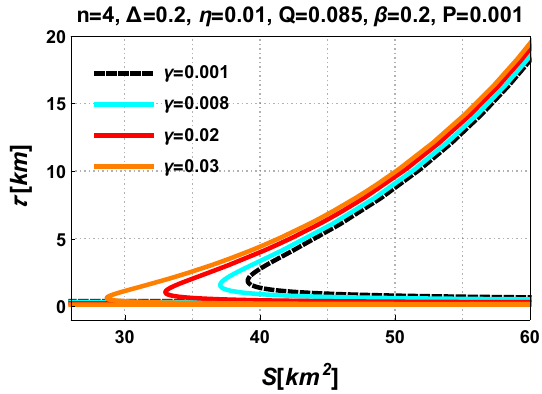}\hspace{1cm}
       \includegraphics[height=6.8cm,width=7.5cm]{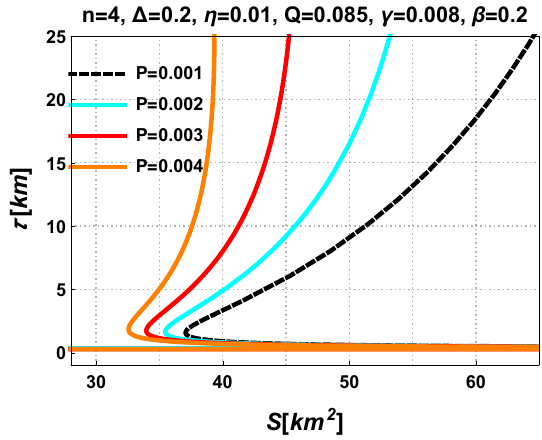}
       \hspace{1cm}
       \includegraphics[height=6.8cm,width=7.5cm]{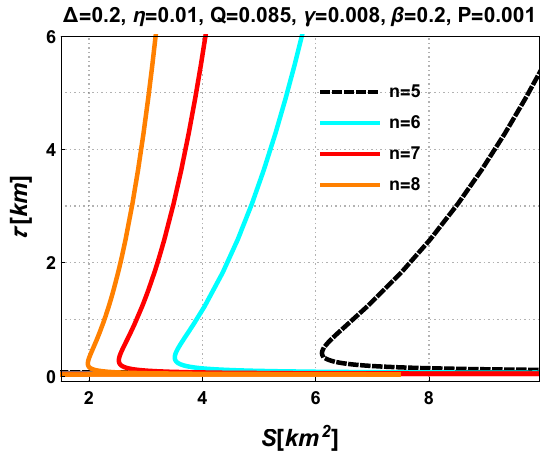}
      }
       
      	\caption{Defect curve $\tau$ versus $S$ for various values of the pressure $P$, the PYM charge $Q$, the PYM index $\Gamma$, the global monopole coupling $\eta$, the Murnaghan parameter $\gamma$ and the spacetime dimensions $n$.} 
    \label{tau}
      \end{figure*}

\begin{figure}[t]
\begin{center}
\subfigure[~Defect curve]  {\label{Parameter_Space_RN_AdSd}
\includegraphics[height=6cm,width=5.5cm]{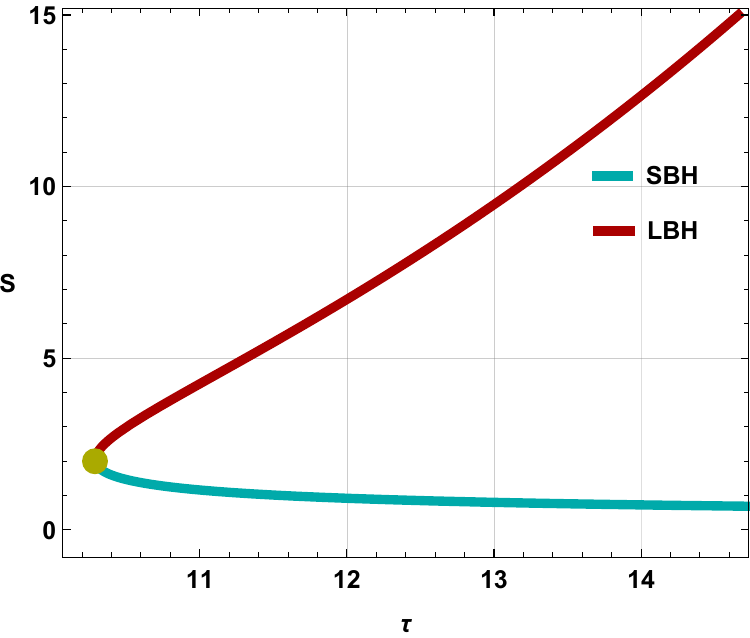}}\hfill
\subfigure[~Vector field]  {\label{Parameter_Space_RN_dSe}
\includegraphics[height=6cm,width=5.5cm]{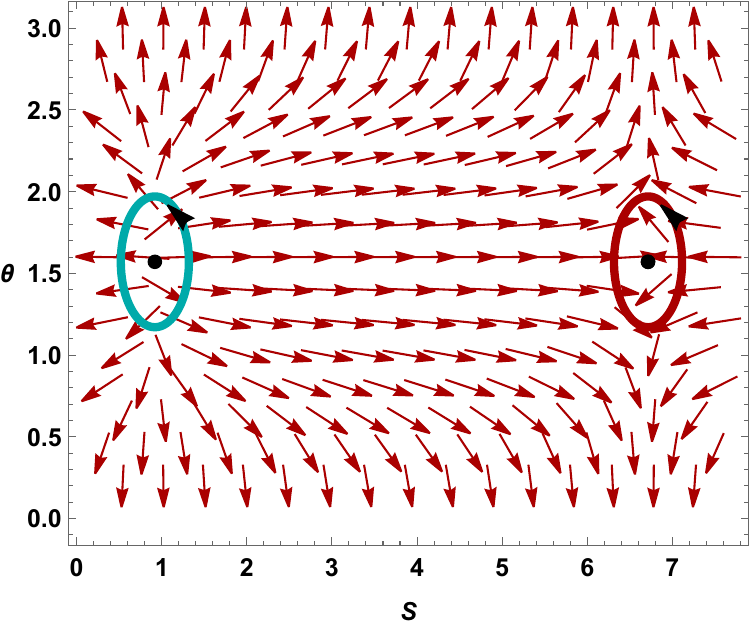} }\hfill
\subfigure[~Winding number]  {\label{Parameter_Space_RN_dSf}
\includegraphics[height=6cm,width=5.5cm]{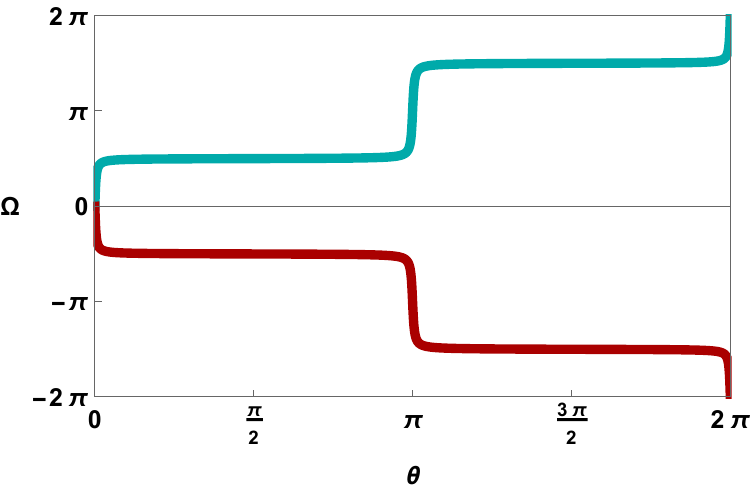}}
\end{center}
\caption{Thermodynamic topology in \textit{five}-dimensional spacetime using $\Delta=0.2,\epsilon=0.95,Q=0.85,\eta=0.01,P=0.001,\beta=0.2,\gamma=0.008,\xi=2,\chi=1$.}
\label{fig13}
\end{figure}

\begin{figure}[t]
\begin{center}
\subfigure[~Defect curve]  {\label{Parameter_Space_RN_AdSd-1}
\includegraphics[height=5.5cm,width=7cm]{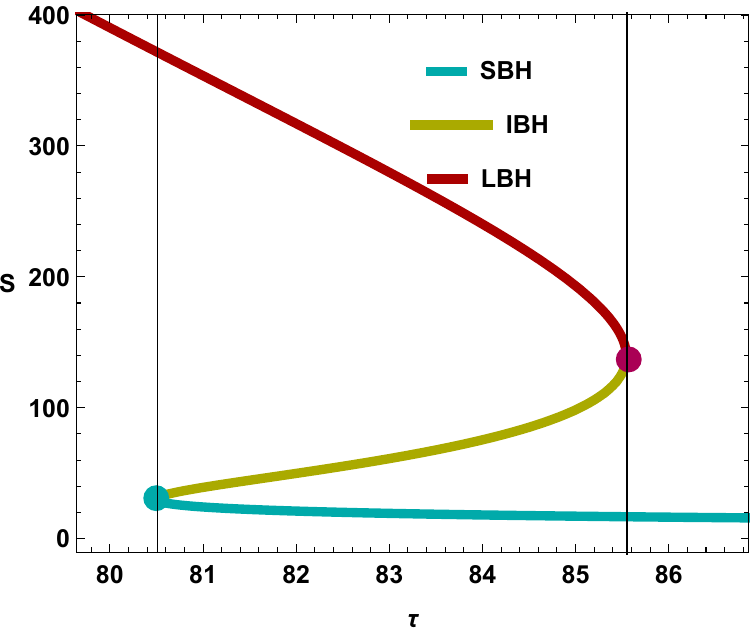}}\hfill
\subfigure[~Vector field]  {\label{Parameter_Space_RN_dSe-1}
\includegraphics[height=5.5cm,width=5.5cm]{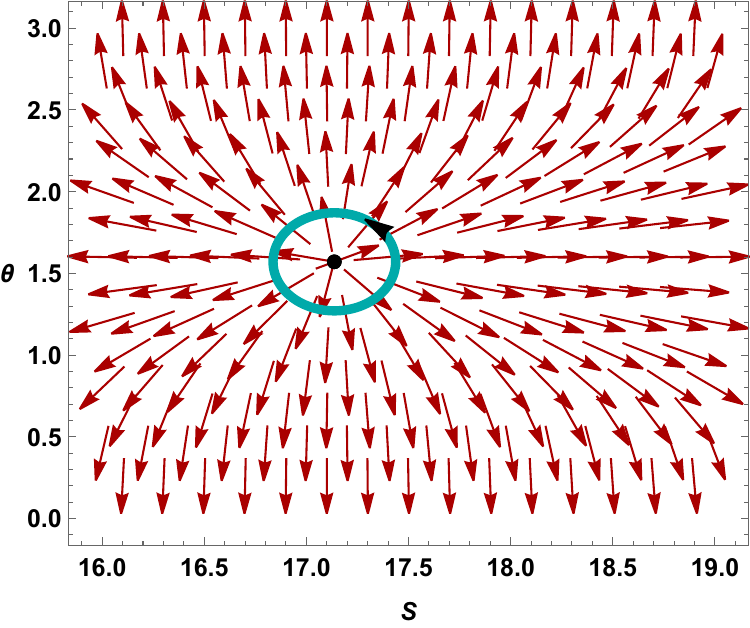} \hspace{1mm}
       \includegraphics[height=5.5cm,width=5.5cm]{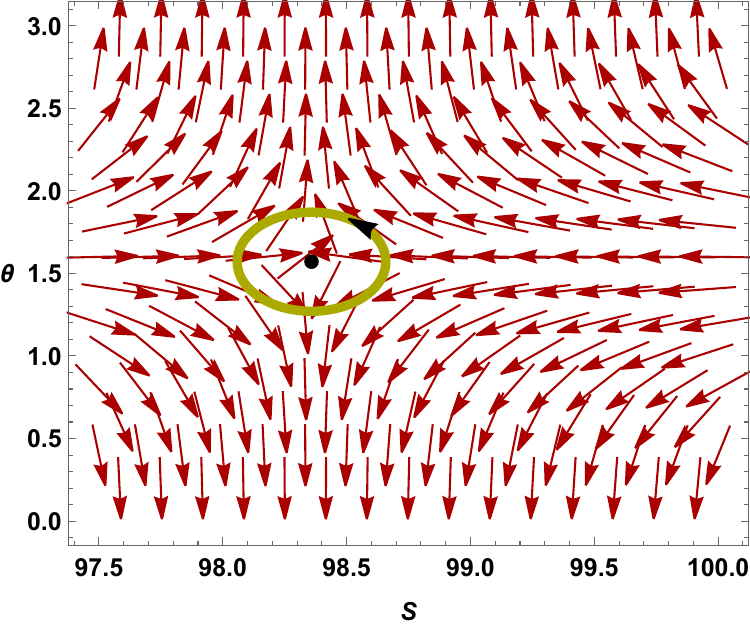}
       \hspace{1mm}
       \includegraphics[height=5.5cm,width=5.5cm]{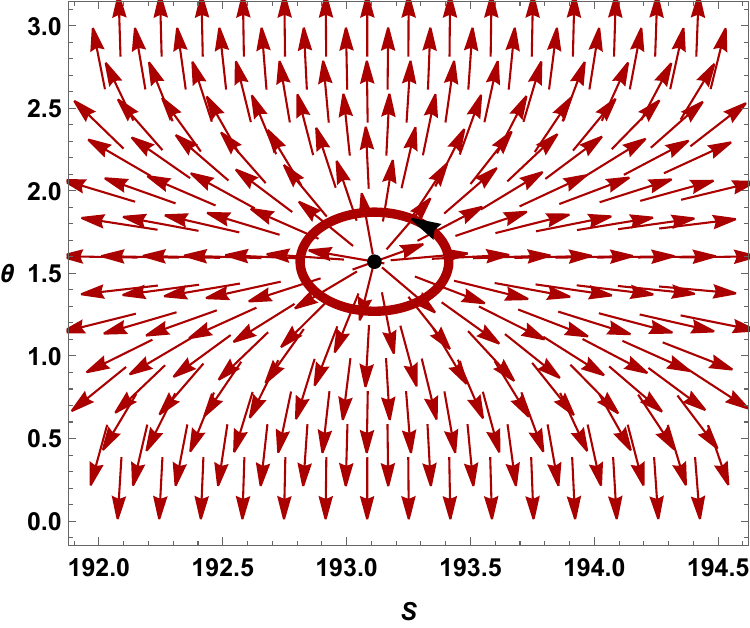}}\hfill
\subfigure[~Winding number]  {\label{Parameter_Space_RN_dSf-1}
\includegraphics[height=6.5cm,width=7.5cm]{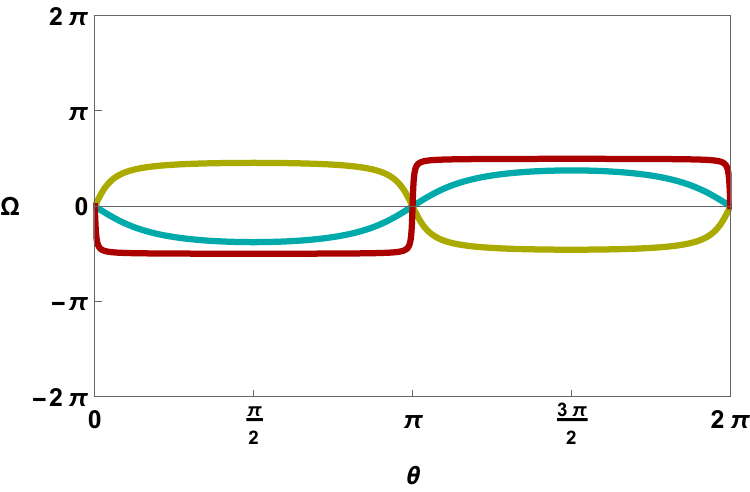}}
\end{center}
\caption{Thermodynamic topology in \textit{four}-dimensional spacetime using $\Delta=0.2,\epsilon=0.95,Q=0.85,\eta=0.01,P=0.001,\beta=0.2,\gamma=0.008,\xi=2,\chi=1$.}
\label{fig14}
\end{figure}
In this formulation, one constructs a conserved topological current in the three-dimensional parameter space $x^\mu=\{t,S,\Theta\}$ \cite{Duan:1998it,Wei2019}:
\begin{equation}
    j^\mu \;=\;\frac{1}{2\pi}\,\varepsilon^{\mu\nu\rho}\,\varepsilon_{ab}\,\partial_\nu n^a\,\partial_\rho n^b,
\end{equation}
where $\varepsilon^{\mu\nu\rho}$ is the Levi–Civita symbol in $(t,S,\Theta)$ and $n^a=\phi^a/\|\phi\|$ is the unit-normalized Duan--$\phi$ field. Conservation, $\partial_\mu j^\mu=0$, follows immediately from the antisymmetry of $\varepsilon^{\mu\nu\rho}$ and $\varepsilon_{ab}$.

Using the Laplacian Green identity in the internal $\phi^a$-space,
$\Delta_{\!\phi}\ln\|\phi\|=2\pi\,\delta^{(2)}(\phi)$, the current may be rewritten as \cite{Xu:2020}:
\begin{equation}
    j^\mu \;=\; \delta^{(2)}(\phi)\;J^\mu\!\biggl(\frac{\phi}{x}\biggr),
\qquad
\varepsilon^{ab}J^\mu\!\biggl(\frac{\phi}{x}\biggr)
=\varepsilon^{\mu\nu\rho}\,\partial_\nu\phi^a\,\partial_\rho\phi^b,
\end{equation}
which makes manifest that $j^\mu$ is nonzero only at the zeros (defects) of $\phi^a$, i.e., at the critical points of the BH free-energy landscape.

The total topological charge $W$ is obtained by integrating the temporal component of the current over a spatial slice $\Sigma$ that encloses the defects \cite{Wei2019}:
\begin{equation}
W=\int_\Sigma j^0\,\mathrm{d}^2x=\sum_{i=1}^N w_i,
\end{equation}
where each winding number $w_i$ may be written in terms of the solid angle (or phase) swept by the unit vector $n^a$ around the $i$-th defect,
\begin{equation}
    w_i \;=\;\frac{1}{2\pi}\,\Omega_i(2\pi),
\qquad
\Omega_i(\nu)\;=\;\int_0^\nu \varepsilon_{ab}\,n^a\,\frac{\mathrm{d}n^b}{\mathrm{d}\nu'}\,\mathrm{d}\nu'.
\end{equation}

In practice we integrate $j^0$ over a compact domain $\Sigma$ in the $(S,\Theta)$ plane bounded by a smooth contour parameterized as
\begin{equation}
S(\nu)=S_0+S_1\cos\nu,\qquad
\Theta(\nu)=\frac{\pi}{2}+S_2\sin\nu,\qquad \nu\in[0,2\pi),
\end{equation}
where $S_0,S_1,S_2$ fix the center and size of the contour. In the absence of zeros of $\phi$ inside $\Sigma$ one has $j^\mu\equiv0$ and hence $W=0$.

With the foregoing background in hand, we now examine the thermodynamic topology of the Murnaghan-PYM system with a GM in the extended parameter space that includes the Barrow deformation $\Delta$. In particular, working in the setting of higher-dimensional AdS BHs and combining the definition \eqref{offshell} with the mass relation \eqref{mass} and the entropy relation \eqref{rbar}, we obtain the off-shell free energy as a function of parameter space with the Barrow parameter $\Delta$ and $\tau$:
\begin{align}\label{off}
\mathcal{F} &= - \frac{S}{\tau}-\frac{1}{8\,\Gamma\!\bigl(\tfrac{n-1}{2}\bigr)}\Biggl\{ 
(n-2)\,\pi^{\frac{n-3}{2}}
\Bigl[\,2^{\frac{1}{n-2}}
\bigl(\pi^{\tfrac12-\tfrac n2}\,\Gamma\!\bigl(\tfrac{n-1}{2}\bigr)\,S^{\tfrac{2}{\Delta+2}}\bigr)^{\!\tfrac{1}{n-2}}
\Bigr]^{n-3}
\notag\\
&\quad\times
\biggl\{
\frac{2^{\tfrac{n}{n-2}}
\bigl(\tfrac{1-\beta}{\gamma}\bigr)^{\!\xi}
\bigl(\pi^{\tfrac12-\tfrac n2}\,\Gamma\!\bigl(\tfrac{n-1}{2}\bigr)\,S^{\tfrac{2}{\Delta+2}}\bigr)^{\!\tfrac{2}{n-2}}
\,_2F_1\!\Bigl(
\xi,\tfrac{\xi}{1-\beta};\tfrac{\xi}{1-\beta}+1;
-\,\frac{\chi^2\!\bigl[\,2^{\tfrac{1}{n-2}}
(\pi^{\tfrac12-\tfrac n2}\Gamma(\tfrac{n-1}{2})S^{\tfrac{2}{\Delta+2}})^{\!\tfrac{1}{n-2}}\bigr]^{-\!\tfrac{(n-1)(\beta-1)}{\xi}}}{\gamma}
\Bigr)}{(n-2)(n-1)}
+\frac{8\pi\,\eta^2}{n-3}
\biggr\}
\notag\\
&\quad
+2^{\tfrac{2}{n-2}}
\bigl(\pi^{\tfrac12-\tfrac n2}\,\Gamma\!\bigl(\tfrac{n-1}{2}\bigr)\,S^{\tfrac{2}{\Delta+2}}\bigr)^{\!\tfrac{2}{n-2}}
\Biggl[
\frac{2\,\Lambda}{n-1}
+\frac{\displaystyle\Bigl(
2^{3-\tfrac{4}{n-2}}
\pi^{3-n}Q^2\Gamma(\tfrac{n-1}{2})^2
\,(\pi^{\tfrac12-\tfrac n2}\Gamma(\tfrac{n-1}{2})S^{\tfrac{2}{\Delta+2}})^{-\!\tfrac{4}{n-2}}
\Bigr)^{\!\epsilon}}
{n-4\epsilon-1}
\Biggr]
\Biggr\}\,. 
\end{align}
This off-shell free energy decomposes into five physically distinct contributions: The term $-S/\tau$ coupling the entropy to the cavity temperature generates the characteristic swallowtail; the overall prefactor $(n-2)\pi^{(n-3)/2}/\bigl[8\,\Gamma((n-1)/2)\bigr]$ together with the $S^{2(n-3)/[(n-2)(\Delta+2)]} $ power encodes the spacetime dimension $n$ and the Barrow entropy deformation $\Delta$; the hypergeometric $_2F_1$ piece arises from integrating the Murnaghan scalar gas EoS (with parameters $(\beta,\xi,\chi,\gamma)$ and controls the depth of the intermediate branch; the constant shift $+8\pi\,\eta^2/(n-3) $ reflects the GM charge $\eta$ and can suppress multivaluedness when large; the term proportional to $S^{2/(n-2)} \left [ \frac{2\Lambda}{n-1} + \frac{(Q^2 \cdots)^{\epsilon}} {n - 4\epsilon - 1} \right] $ encapsulates the AdS (or dS) cosmological constant and the YM charge $Q$ with exponent $\epsilon$. These elements are responsible for seeding the large and small BH wells, respectively. In the small-$S$ limit, the scalar gas and charge terms drive $\mathcal{F}\to-\infty$, creating the small-BH branch, while for large $S$, the competition between $-S/\tau$ and $\Lambda\, S^{2/(n-2)}$ yields the large-BH branch. Three real extrema, as observed in $n=4$, lead to the formation of a swallowtail and facilitate a first-order phase transition. In contrast, when $n=5$, the milder charge exponent $(Y \sim S^{-1.212})$ inhibits the occurrence of three extrema, resulting in a smooth crossover instead. This decomposition elucidates how the parameters $\Delta, \beta, \xi, \eta, Q, \epsilon, \Lambda$ collectively shape the thermodynamic landscape and influence the existence or non-existence of a first-order coexistence region.

Therefore, using Eq.~\eqref{phi}, the components of the vector $\phi$ can be determined as
\begin{align}
\phi^{r}(S)
&= -\frac{1}{\tau}
+ \mathcal{K}(S;\Delta,n)\,
  \Lambda(S;\beta,\xi,\chi)\,
  \Sigma(S;Q,\epsilon,\eta)\,, 
\\[1cm]
\phi^{\Theta}(\Theta)
&= -\cot \Theta\,\csc \Theta\,,
\end{align}

where

\[
\begin{aligned}
\mathcal{K}(S;\Delta,n)
&= \frac{1}{(n-2)\,\pi\,(2+\Delta)}
  \,2^{-2 - \frac{3}{n-2}}
  \,S^{-\frac{\Delta}{2+\Delta}}
  \Bigl[\,
    \pi^{\tfrac12-\tfrac n2}\,
    \Gamma\!\bigl(\tfrac{n-1}{2}\bigr)\,
    S^{\tfrac{2}{2+\Delta}}
  \Bigr]^{\frac{1+n}{2-n}}, 
\\[0.5em]
\Lambda(S;\beta,\xi,\chi)
&= \Bigl[
    1 + 
    \frac{\chi^2\,
      \bigl(
        2^{\tfrac{1}{n-2}}
        \bigl(\pi^{\tfrac12-\tfrac n2}\,\Gamma(\tfrac{n-1}{2})\,S^{\tfrac{2}{2+\Delta}}\bigr)^{\!\tfrac{1}{n-2}}
      \bigr)^{-\frac{(n-1)(\beta-1)}{\xi}}}
    {\gamma}
  \Bigr]^{-\xi}, 
\\[0.5em]
\Sigma(S;Q,\epsilon,\eta)
&= -2^{\tfrac{n}{n-2}}
   \Bigl(\frac{1-\beta}{\gamma}\Bigr)^{\!\xi}
   \bigl(\pi^{\tfrac12-\tfrac n2}\,\Gamma(\tfrac{n-1}{2})\,S^{\tfrac{2}{2+\Delta}}\bigr)^{\tfrac{2}{n-2}}
  +6\,
  \Bigl[
    1 + 
    \frac{\chi^2\,
      \bigl(
        2^{\tfrac{1}{n-2}}
        \bigl(\pi^{\tfrac12-\tfrac n2}\,\Gamma(\tfrac{n-1}{2})\,S^{\tfrac{2}{2+\Delta}}\bigr)^{\!\tfrac{1}{n-2}}
      \bigr)^{-\frac{(n-1)(\beta-1)}{\xi}}}
    {\gamma}
  \Bigr]^{\!\xi}
\\[-0.3em]
&\quad
  + 
  \Bigl[
    1 + 
    \frac{\chi^2\,
      \bigl(
        2^{\tfrac{1}{n-2}}
        \bigl(\pi^{\tfrac12-\tfrac n2}\,\Gamma(\tfrac{n-1}{2})\,S^{\tfrac{2}{2+\Delta}}\bigr)^{\!\tfrac{1}{n-2}}
      \bigr)^{-\frac{(n-1)(\beta-1)}{\xi}}}
    {\gamma}
  \Bigr]^{\!\xi}
  \biggl(
    n^2 + 16\pi\,\eta^2 - n\,(5+8\pi\,\eta^2)
\\[-0.3em]
&\quad\quad
    -\,2^{\tfrac{2}{n-2}}
    \bigl(\pi^{\tfrac12-\tfrac n2}\Gamma(\tfrac{n-1}{2})S^{\tfrac{2}{2+\Delta}}\bigr)^{\tfrac{2}{n-2}}
    \Bigl[
      -16\,P\,\pi
      + 
      \bigl(
        2^{3-\tfrac{4}{n-2}}
        \pi^{3-n}\,
        Q^2\,
        \Gamma\!\bigl(\tfrac{n-1}{2}\bigr)^2\,
        (\pi^{\tfrac12-\tfrac n2}\Gamma(\tfrac{n-1}{2})S^{\tfrac{2}{2+\Delta}})^{-\tfrac{4}{n-2}}
      \bigr)^{\!\epsilon}
    \Bigr]
  \biggr).
\end{aligned}
\]
Within Duan’s topological framework,  $\mathcal{K}(S)$ collects all Barrow entropy and dimensional prefactors, $\Lambda(S)$ encodes the full Murnaghan hypergeometric structure, whose derivative generates the non-trivial “wiggle” in $\phi^{r}(S)$, and $\Sigma(S)$ sums the YM-gas, monopole, and charge correction monomials.  Zeros of $\phi$ satisfy  
$\phi^r=0$ at $\Theta=\pi/2$,  
and the Jacobian $\det(\partial_i\phi^j)|_{(S_c,\pi/2)}=\phi^{r}{}'(S_c)\times(-2)$ confirms each is a nondegenerate Duan critical point (Morse index $\pm1$). On the other hand, in Duan’s picture, each zero of $\phi$ labels a distinct thermodynamic branch: small‐$S$ roots stem from the negative YM-gas power, medium‐$S$ roots from the scalar gas hypergeometric term, and large-$S$ roots (if any) from the monopole/cosmological sector.  Raising $\xi$ (at $\beta<1$) deepens the medium‐$S$ well in $\phi^r$, expanding the metastable window, while increasing $\eta$ uniformly elevates $\phi^r$, merging multiple Duan zeros into a single stable point and thereby suppressing first-order transitions.  Thus, $(\Delta,\beta,\xi,\eta)$ together shape the topological landscape of BH phases in both four and five dimensions.

We can now identify the zero points of the vector field. One such zero point consistently occurs at $\Theta = \frac{\pi}{2}$, due to the parameterization~\eqref{phi} of the $\Theta$-component of the field. Additionally, 
we derive an equation for $\tau$ by solving $\phi^{r_+} = 0$, which gives
\begin{equation}
\tau = \frac{%
2^{\,2 + \tfrac{3}{n-2}}\,(n-2)\,\pi\,(2+\Delta)\,
S^{\tfrac{\Delta}{2+\Delta}}\,
A^{\!\tfrac{1+n}{n-2}}\,
B^{-n}\,(1+X)^{\xi}
}{%
-\,2^{\tfrac{n}{n-2}}\Bigl(\tfrac{1-\beta}{\gamma}\Bigr)^{\xi}A^{\!\tfrac{2}{n-2}}
\;+\;6\,(1+X)^{\xi}
\;+\;(n^2+16\pi\eta^2 -n\,(5+8\pi\eta^2))\,(1+X)^{\xi}
\;-\;2^{\tfrac{2}{n-2}}\,A^{\!\tfrac{2}{n-2}}\,
\bigl[-16P\pi + Y\bigr]
}\,,
\end{equation}
with 
\begin{align}
    A = \pi^{\tfrac12-\tfrac n2}\,\Gamma\!\bigl(\tfrac{n-1}{2}\bigr)\,S^{\tfrac{2}{2+\Delta}},
\,\quad
B = 2^{\tfrac{1}{n-2}}\,A^{\tfrac{1}{n-2}}, 
\,\quad
X = \frac{\chi^2\,B^{-\tfrac{(n-1)(\beta-1)}{\xi}}}{\gamma},
\,\quad
Y = \Bigl(2^{\,3-\tfrac{4}{n-2}}\,\pi^{3-n}\,Q^2\,\Gamma\!\bigl(\tfrac{n-1}{2}\bigr)^2\,A^{-\tfrac{4}{n-2}}\Bigr)^{\!\epsilon}.
\nonumber
\end{align}
Within this closed-form EoS the functions $A$, $B$, $X$ and $Y$ serve to streamline the dependence on entropy $S$ and parameters $(\Delta,\beta,\xi,\chi,\epsilon,Q,\eta,P)$. The overall prefactor in the numerator where $\;2^{2+\tfrac{3}{n-2}}(n-2)\,\pi\,(2+\Delta)\, S^{\tfrac{\Delta}{2+\Delta}} A^{\tfrac{1+n}{n-2}}\,B^{-n}(1+X)^{\xi}$, encodes the Barrow entropy deformation as $S^{\Delta/(2+\Delta)}$, the dimensional scaling $n$, and the scalar gas amplitude $(1+X)^{\xi}$. In the denominator, the term proportional to $\bigl(\tfrac{1-\beta}{\gamma}\bigr)^{\!\xi}A^{2/(n-2)}$ arises from the Murnaghan backreaction, the factor $6(1+X)^{\xi}$ from the pure scalar gas pressure, and the combination $(n^2+16\pi\eta^2-n(5+8\pi\eta^2))(1+X)^{\xi}$ from the global-monopole contribution. Additionally, the element with $Y$ and the pressure $P$ captures the non-linear YM charge correction. Altogether, this structure makes manifest how tuning $\beta,\xi$ adjusts the relative weight of hypergeometric versus scalar gas branches, while $\eta$ enters as a uniform shift and $Q,\epsilon$ drive the small entropy behavior through $Y$.

Figure \ref{tau} shows the variation of the defect curve $\tau(S)$ for various values of the BH parameter space together with the $\Delta$-Barrow parameter. It is observed that when $\Delta$ rises from the Bekenstein-Hawking limit to range over Barrow space ($0\to0.3$), the $\tau(S)$ curve develops a significant “wiggle” with two real roots of $\phi^r=0$. These two zeros bracket a metastable branch: the smaller-$S$ root indicates the small/intermediate coexistence point, and the larger-$S$ root the intermediate/large phase transition. The interpretation of this phenomenon suggests that a larger value of $\Delta$ increases the sharpness of the swallowtail bifurcation, thereby increasing the range of the coexistence window. Increasing $\eta$ uniformly raises $\tau(S)$, ultimately leading to the merging and disappearance of the two zeros. Above a critical value, $\eta_{\rm crit}$, only a single root remains, resulting in the complete suppression of first-order transitions. On the other hand, varying the YM charge $Q$ or the Murnaghan parameters $\gamma$ and $\beta$ shifts the location and depth of the wiggle. Moderate values of $Q$ or $\gamma$ bring the roots closer together, thereby altering the critical temperatures associated with the transitions between small/intermediate and large states. Additionally, for $n = 4$, $\tau(S)$ consistently exhibits two zeros for the specified parameters, indicating a small–large phase state with a total topological charge of $W = +1$. In contrast, for $n \geq 5$, the curve is monotonic, resulting in the survival of only one zero (the large-$S$ branch), and no bifurcation occurs, yielding a net $W = 0$.

From a graphical perspective, Figures \ref{fig13}-\ref{fig14} illustrate the defect curve $\tau(r_h)$, the unit vector field $n$, and the phase winding $\Omega(\theta)$ for $n=4$ and $n =5$, using the benchmark set $\{\Delta=0.2,\epsilon=0.95,Q=0.85,\eta=0.01,P=0.001,\beta=0.2,\gamma=0.008,\xi=2,\chi=1\}$.  In the four-dimensional spacetime ($n=4$) scenario, the charge correction term scales in asymptotic expansion as $Y\sim S^{-1.727}$, and thus dominates the denominator $\text{Denom}\,\tau(S)$ for $S\ll1$. This causes $\text{Denom}\,\tau(S)$ to reach zero at a critical entropy $S_c^{(4)}\approx0.2165$. Consequently, a vertical asymptote in $\tau(S)$ could occur, which indicates a narrow first-order phase transition similar to the Van der Waals transition. In other words, the four-dimensional case ($n=4$) is characterised by the existence of three BH branches: small, intermediate, and large ones. Upon these observations, one can draw the findings into the so-called winding number process, illustrating the zero point of $\phi^{S} = 0$. Indeed, it is found in the realm of the four-dimensional case $(n=4)$ that the zero point of $\phi^{S}$ exhibits three roots (three contours (S/I/L) $C_i$, i.e., two critical points $Cp_i$). Conventionally and coherently with our proper observations, for the small and large BH, the winding number at the zero points is $w=+1$ (stable phase), while for the intermediate BH, the winding number is $w=-1$ (unstable phase). As a result, the total topological charge is set to $W=2\times(+1)-1=1$. Graphically, the presence of two critical points underlying three contours with different colours are located at $(S,\theta) = (0.38871,\pi/2)$, $(S,\theta) = (3.71934,\pi/2)$ and $(S,\theta) = (5.26963,\pi/2)$. In addition, the phase winding $\Omega(\theta)$ involving three contours is depicted at the bottom of Figure \ref{fig14}, demonstrating that the total topological charge is fixed at $W=1$. The phenomenon of a triple root (zeros of $\phi^{S} = 0$) is intriguing; for this to be achieved, two of these critical points (critical points $CP_1$ (red contour) and $CP_2$ (magenta contour)) are endowed with a physical structure. However, the third critical point (represented by the blue contour) is unphysical. Thus, it cannot minimize the Gibbs free energy, i.e., it represents a mathematical rather than thermodynamic critical point.

In the five-dimensional case $(n=5)$ as illustrated in Figure \ref{fig13}, the exponent governing the $Y$-term becomes milder, scaling as $Y \sim S^{-1.212}$. Consequently, the defect curve $\tau(S)$ admits exactly two real zeros of $\phi_{r}(S)=0$, which correspond to the single extremum of $\tau(S)$. The angular field $\Theta(S)$ then exhibits one local maximum--associated with the unstable, small--BH branch--and one local minimum--associated with the stable, large--BH branch. Calculating the phase winding $\Omega(\Theta)$ around these two critical points yields winding numbers
\[
  w = -1 \quad\text{(small-$S$ defect)}, 
  \qquad
  w = +1 \quad\text{(large-$S$ defect)},
\]
which sum up to a net topological charge $W = (-1) + (+1) = 0$. A vanishing total charge indicates the disappearance of a genuine first-order small/large coexisting phase; instead, the system undergoes a smooth crossover in its thermodynamic behavior.

	



\section{Conclusion}\label{sec5}

We have derived and analysed an exact family of higher-dimensional BH solutions in Einstein-–PYM gravity minimally coupled to a GM and a surrounding scalar gas governed by a generalized Murnaghan (dual-polytropic) EoS.  The analytic construction obtained by the WY magnetic ansatz and the radial condition $p_{r}=-\rho$ yields a lapse function $F(r)$ expressible in closed form (elementary functions plus Gauss hypergeometric contributions) and an explicit scalar density profile. This exactness is the first and most important practical achievement of the present work: it provides a controlled, tractable laboratory in which to trace how gauge nonlinearity, topological defects and non-linear matter jointly backreact on horizon structure, curvature invariants and thermodynamic observables.  Concretely, the closed-form nature of the solution enabled direct evaluation of curvature scalars, verification of EC domains, and high-precision numerical representation of thermodynamic functions (Hawking temperature $T(r_h)$, heat capacity $C_{P,Q}$, and associated topological maps).  These results show that the model is not qualitatively merely interesting but also quantitatively predictive across a wide parameter space.

The geometric and local-physics analysis reveals a structured and parameter-sensitive spacetime manifold.  Curvature invariants indicate a classical central singularity for $n>3$ in the parameter ranges explored, while asymptotics are dictated by the cosmological constant $\Lambda$. ECs testing demonstrates that the matter configuration is generically SEC-violating (phantom-like) but may satisfy NEC and DEC in substantial exterior regions for physically reasonable choices of $\beta,\gamma,\chi$.  Dimension $n$ plays an essential role in moderating small--scale behavior: higher dimensions dilute local curvature contributions (via the entropy scaling $S\propto r_h^{\,n-2}$) and tend to soften heat-capacity singularities.  Taken together, these findings delineate the physically acceptable windows of the model and indicate where classical regularization (or quantum corrections) will be required if one desires a globally regular solution.

The thermodynamic and topological analysis uncovers a richly structured phase landscape.  The Hawking temperature $T(r_h)$ is generically non-monotonic and exhibits multiple extrema whose radii coincide with changes in thermodynamic stability.  The heat capacity $C_{P,Q}(r_h)$ shows zeros and poles that delimit stable and unstable branches; poles correspond to continuous (second-order) critical points, while sign changes in $C_{P,Q}$ signal local thermodynamic instabilities.  By mapping the thermodynamic SF into the Duan $\phi$-map we identified topological changes in winding number that accompany branch creation/annihilation, establishing that the system admits genuine thermodynamic topology transitions as parameters vary.  These results are robust across the tested dimensionalities and parameter slices and are directly tied to physical observables: poles and swallowtails in the free energy (suggested by the $T$ and $C$ behavior) indicate a coexistence regimes that can, in principle, be classified by standard Maxwell constructions.

A central, actionable output of the paper is a detailed account of how each model parameter sculpts the geometry and the thermodynamics.  Summarised succinctly: (i) the PYM charge $Q$ is the primary scale-setter  increasing $Q$ shifts divergent points and extrema to larger $r_h$ and amplifies transition strength, effectively stabilising larger BHs; (ii) the PYM nonlinearity index $\epsilon$ intensifies short-scale curvature, producing tall, narrow singularities in $C_{P,Q}$ and thereby enhancing small-radius criticality; (iii) the global-monopole parameter $\eta$ acts predominantly as a geometric translation (solid-angle deficit) that displaces thermodynamic features collectively without introducing new small-scale singularities; (iv) the Murnaghan barotropic coefficient $\beta$ controls the crossover between core and asymptotic behavior  larger $\beta$ tends to hide intermediate transitions by pushing Murnaghan features to smaller radii, producing degeneracies with other parameters; (v) the Murnaghan modulus $\gamma$ quantifies scalar-backreaction strength — small $\gamma$ (strong backreaction) creates additional intermediate divergent points and complexifies the phase diagram independent of the gauge sector; and (vi) the spacetime dimension $n$ and the cosmological constant $\Lambda$ modulate the quantitative manifestation of all preceding effects (dimensional dilution and asymptotic thermodynamic backdrop respectively).  This parameter taxonomy not only clarifies the origin of the multi-branch topology visible in Figures \ref{fig5a}--\ref{fig6f}, but also provides a practical roadmap for selecting physically admissible parameter sets for further phenomenological study.

Although our PYM-GM-Murnaghan family already exhibits rich structure, several interrelated challenges deserve to be focused in a future study. First, the construction and dynamical evolution of fully rotating PYM-GM-Murnaghan BH and, in particular, the control of horizon stability demand state of the art numerical relativity techniques and tailored stability diagnostics. Secondly, the intrinsically non-linear coupling between the PYM sector and the GM core, together with a back-reacting Murnaghan fluid, may qualitatively modify causal structure, excite new instability channels or support long-lived quasi-stationary or soliton-like configurations; these possibilities require both perturbative classification and non-linear evolution studies. Third, the Murnaghan EoS changes the effective stress-energy profile and thus is expected to leave an imprint on linear perturbations, quasi-normal modes (QNMs) spectrum and possible mode mixing, so a systematic computation of QNMs and their observational signatures is necessary. Fourth, monopole and fluid contributions can alter the thermodynamic phase structure and topological invariants (including possible reentrant or multimodal behavior), so mapping the full phase diagram and including semi-classical corrections (one-loop determinants, generalised entropy terms) is essential to connect with holographic and semi-classical methods. Finally, confronting theory with observation requires quantifying empirical discriminants modified ringdown templates, gravitational wave echoes, and deviations in shadow/lensing observables that could distinguish PYM-GM-Murnaghan configurations from standard solutions.

\section*{Acknowledgments}
The author SKM is also thankful to the UoN administration for the continuous support and encouragement for the research work.


\end{document}